\documentclass[aps,prd,showpacs,notitlepage,nofootinbib,superscriptaddress,floatfix,showkeys,twocolumn]{revtex4-1}

\usepackage[T1]{fontenc} 

\usepackage[dvipsnames]{xcolor}
\usepackage{amsmath}
\usepackage{amssymb}
\usepackage{mathtools}
\usepackage{physics}

\usepackage{enumerate}
\usepackage{feynmp-auto}
\usepackage{float}
\usepackage{subcaption}
\usepackage[utf8]{inputenc}
\usepackage{ragged2e}

\usepackage[normalem]{ulem}

\usepackage{hyperref}
\usepackage[all]{hypcap}
\hypersetup{  
    colorlinks=true,
    linkcolor=blue,
    filecolor=red,      
    urlcolor=cyan,
    citecolor=green,
    }

\usepackage{slashed}

\newcommand{\tdom}{T_{\rm dom}}

\newcommand{\tdec}{T_{\rm dec}}

\newcommand{\nt}{n_T}
\newcommand{\trh}{T_{\rm RH}}
\newcommand{\keq}{k_\text{eq}}
\newcommand{\krh}{k_\text{RH}}
\newcommand{\kdec}{k_\text{dec}}
\newcommand{\kdecs}{k_\text{dec, S}}
\newcommand{\krhs}{k_\text{RH, S}}

\begin{document}

\title{Axion Dark Matter Archaeology with Primordial Gravitational Waves}

\author{Andrew Cheek}
\email{acheek@sjtu.edu.cn}
\affiliation{Tsung-Dao Lee Institute \& School of Physics and Astronomy, Shanghai Jiao Tong University, Shanghai 201210, China}

\author{Anish Ghoshal}
\email{anish.ghoshal@fuw.edu.pl}
\affiliation{\,Institute of Theoretical Physics, Faculty of Physics, University of Warsaw, \\ ul. Pasteura 5, 02-093 Warsaw, Poland}

\author{Debarun Paul}
\email{debarun31paul@gmail.com}
\affiliation{\,Physics and Applied Mathematics Unit, Indian Statistical Institute, \\ 203 B.T. Road, Kolkata 700108, India}

\begin{abstract}
    We investigate the complementary information to be gained from inflationary gravitational wave (IGW) signals and searches for QCD axion dark matter. We focus on post-inflationary Peccei-Quinn (PQ) breaking axion models that are cosmologically safe. Recent work has shown that a greater number of such models exist. This is because the heavy quarks required for the colour anomaly can provoke a period of heavy quark domination (HQD), which, through decay, dilutes the axion abundance. In this work we show for the first time that the axion dark matter mass can be as low as $m_a\sim10^{-8}\,{\rm eV}$ for models where the heavy quarks decay via dimension 6 terms. This is achieved by allowing the mass of the heavy quarks to differ from the axion decay constant, $m_Q\neq f_a$. Consequently, the observables that would distinguish between pre- and post-inflationary PQ breaking, $m_a$ and the additional relativistic degrees of freedom $\Delta N_{\rm eff}$, now become indiscernible. To solve this, we propose using blue-tilted IGWs to probe HQD. In scenarios where such a blue tilt is present, the enhanced GW signal allows future interferometers to place non-trivial constraints on the parameters $m_Q$ and $f_a$, thereby complementing haloscope searches. While some degeneracies with other parameters such as $m_Q$ remain, detectors such as BBO and ET  will be able to optimistically probe $f_a\gtrsim 10^{14}\,{\rm GeV}$.
\end{abstract}

\maketitle

\section{Introduction}
\label{sec:intro}

The QCD axion provides a compelling an elegant explanation of both the strong CP problem and dark matter simultaneously. The axion itself is the proposed pseudo-Nambu-Goldstone boson of a new spontaneously broken $U(1)_{\rm PQ}$ symmetry~\cite{Peccei:1977hh, Peccei:1977np,Weinberg:1977ma, Wilczek:1977pj}. As a consequence the axion's mass, $m_a$, can naturally be very light and can be related to the breaking scale $f_a$ via 
\begin{equation}
    m_a = 5.7\times10^{-5}\,{\rm eV} \left(\frac{10^{11}\,{\rm GeV}}{f_a}\right).
    \label{eq:QCDaxionmass}
\end{equation}
Such a light boson in an expanding Universe will undergo a non-thermal production mechanism known as vacuum misalignment~\cite{Preskill:1982cy, Dine:1982ah, Abbott:1982af} (see also reviews in Refs.~\cite{Sikivie:2006ni,Marsh:2015xka}). This mechanism produces a cold dark matter abundance of axions which can constitute all of the relic dark matter inferred from cosmology and astrophysics~\cite{Cirelli:2024ssz}. This implies a very high symmetry breaking scale, $f_a\sim10^{11}-10^{12}\,{\rm GeV}$ with the corresponding axion mass, $m_a$ being $\mathcal{O}(10)\,\mu{\rm eV}$. Therefore, by searching for evidence of the axion, we will be indirectly probing physics at energy scales far beyond the reach of current particle physics colliders. 

Fortunately, the solution to the CP problem necessitates some interactions between the axion and the Standard Model (SM), the strength of which is also suppressed by $f_a$. Despite this large suppression, axion emission from supernovae~\cite{Carenza:2019pxu} as well as precision measurement of hadron decays at experiments like NA62~\cite{NA62:2017rwk} and KOTO~\cite{Yamanaka:2012yma} provide lower bounds on $f_a$~\cite{MartinCamalich:2020dfe, Alonso-Alvarez:2023wig}. Furthermore, dedicated axion dark matter experiments such as ADMX~\cite{ADMX:2009iij,ADMX:2019uok}, CAPP~\cite{Kim:2023vpo}, UF~\cite{Hagmann:1996qd}, TASEH~\cite{TASEH:2022vvu}, HAYSTAC~\cite{HAYSTAC:2018rwy} and QUAX~\cite{QUAX:2023gop,QUAX:2024fut} have achieved excellent sensitivities, and future experiments like MADMAX~\cite{2019EPJC...79..186B,2024arXiv240802368E}, ORGAN~\cite{McAllister:2017lkb,10.1007/978-3-030-43761-9_5}, FLASH~\cite{Alesini:2017ifp,Alesini:2023qed}, BabyIAXO-RADES~\cite{Ahyoune:2023gfw}, DMRadio~\cite{DMRadio:2022pkf} promise to survey large portions of the parameter range where QCD axion dark matter is predicted. 

By probing energies far beyond the ${\rm TeV}$-scale we are indirectly exploring physics of the very early Universe, an epoch we know very little about~\cite{Allahverdi:2020bys}. Indeed we can only say with certainty that there was a period of cosmic inflation which ended and by the time of Big Bang Nucleosynthesis (BBN), $T\sim 3\, {\rm MeV}$, the Universe was dominated by SM radiation. This provides ambiguity about the dynamics of any high-scale extension to the SM such as the those responsible for the PQ mechanism. For example, the consequences of symmetry breaking above or below the temperature of reheating after inflation are different. If the $U(1)_{\rm PQ}$ symmetry is broken after inflation many axion models predict the formation of stable topological defects such as domain walls~\cite{Zeldovich:1974uw, Sikivie:1982qv}, which require additional mechanisms to destroy~\cite{Barr:2014vva, Reig:2019vqh, Caputo:2019wsd}. Recently, it has been shown that a number of models that do not have a domain wall problem~\cite{DiLuzio:2016sbl, DiLuzio:2017pfr} will generate a period of early heavy quark domination (HQD)~\cite{Cheek:2023fht, DiLuzio:2024xnt} akin to early matter domination (EMD). As a consequence the measurable parameters that would distinguish the pre and post-inflationary symmetry breaking scenarios are equivalent. The task is then to find additional probes on the time of $U(1)_{\rm PQ}$ breaking in relation to inflation. We show that inflationary first-order gravitational waves (GWs), in tandem with axion haloscope experiments will be able do this. We promote such tests of axion dark matter in complementary searches between laboratory and GW experiments. 

In this work we focus on GWs that are generated from quantum tensor fluctuations of the metric during inflation~\cite{Grishchuk:1974ny,Starobinsky:1979ty,Rubakov:1982df,Guzzetti:2016mkm}. The spectrum of such GWs follow a simple power law during radiation domination acting as a tracer of cosmic  expansion ~\cite{Seto:2003kc,Boyle:2005se,Boyle:2007zx,Kuroyanagi:2008ye,Nakayama:2009ce,Kuroyanagi:2013ns,Jinno:2013xqa,Saikawa:2018rcs,Chen:2024roo,Ghoshal:2024gai,Haque_2021,Chakraborty:2023ocr,Maity:2024cpq}. Any departure from the standard thermal history, such as HQD provides distinctive GW spectral features ~\cite{Berbig:2023yyy,Borboruah:2024eha,Borboruah:2024eal,Bernal:2020ywq, Datta:2022tab, Datta:2023vbs, Chianese:2024nyw, Borboruah:2025hai}. Investigating such features  reveals both the time at which early matter domination began and its duration. We will show that this allows us to get a handle on the mass of the particle responsible for HQD, and the rate of their decay. Only when combining this information with axion haloscope experiments can we determine the breaking scale $f_a$. We therefore show that complementarity between GW and axion experiments will be vital for reconstructing the relevant particle physics parameters.

\textit{This paper is organized as follows,} in section~\ref{sec:axion_model} we describe effective axion model framework that we utilize in this article. We discuss how EMD is generated and show how such a scenario changes the expected value for $m_a$. We then discuss how this complicates phenomenological implications for pre and post-inflationary $U(1)_{\rm PQ}$ scenario. In section~\ref{sec:pgw} we review the modelling of inflationary first-order GWs and how they can be used to detect a period of EMD before BBN. We also describe our limit setting methodology which we undertake for all relevant current and future GW detectors. In section \ref{sec:haloscope_GWs} we present our results, combining axion haloscope and GW experiments, we discuss the current status and future potential. Finally we conclude in section~\ref{sec:conclusion}.

\medskip

\section{Preferred Axion models with Heavy Quarks}
\label{sec:axion_model}
The cornerstone of the Peccei-Quinn (PQ) solution~\cite{Peccei:1977hh, Peccei:1977np} to the strong CP (charge parity) problem is to propose a dimension-5 term, the colour anomaly, in addition to the CP violating term allowed by SM gauge invariance. These enter the Lagrangian as
\begin{equation}
    \mathcal{L}_{\text{PQ}} \supset  N \frac{a\left(t, x\right)}{f_a} \frac{g_s^2}{32\pi^2} \tilde{G}^{\mu\nu}_b G_{\mu\nu}+\theta \frac{g_s^2}{32\pi^2} \tilde{G}^{\mu\nu}_b G_{\mu\nu},\label{eq:LCP}
\end{equation}
where $a$ is the axion field, $G$ and $\tilde{G}$ are the field strength tensors of the gluon field and its dual respectively. The coefficients $g_s$, $\theta$, $N$, and $f_a$ are the SM strong coupling constant, the QCD vacuum angle, a model-dependent anomaly coefficient, and the axion decay constant respectively. The first term, under the influence of the QCD potential is dynamically driven to cancel the second term, resulting in no detectable CP violation from strong interactions. How this dimension-5 operator is generated is a model building exercise. Typically there are two approaches: either with SM particles charged under the new $U(1)_{\rm PQ}$ (DSFZ-type~\cite{Zhitnitsky:1980tq,Dine:1981rt,Kim:2008hd}) or not (KSVZ-type~\cite{Kim:1979if,Shifman:1979if}). In both cases new particles are required, for KSVZ, the minimal requirement is to introduce two new fields: a new $SU(3)_{c}$ charged fermion, $Q$, a heavy quark; and a complex scalar field $\Phi$ which drives the PQ breaking. The PQ Lagrangian is given by
\begin{equation}
    \mathcal{L}_{\rm PQ}= \vert\partial_\mu\Phi\vert^2 +\overline{Q}i\slashed{D}Q - (y_Q\overline{Q}_L Q_R \Phi + {\rm H.c.})\,.
    \label{eq:LagPQ}
\end{equation}
Where $y_Q$ is the BSM Yukawa coupling which generates Q mass when PQ symmetry is broken. The PQ charge of $\Phi$ is $1$ and the chiral left-handed $Q_L$ and right-handed $Q_R$ fields have different charges $\chi_L$ and $\chi_R$ respectively, but with $\vert\chi_L-\chi_R\vert=1$ such that the above Lagrangian is symmetric under $U(1)_{\rm PQ}$. In the broken phase, the Yukawa term reduces to  a mass term with $m_Q=y_Q f_a/\sqrt{2}$. If the breaking scale is below the reheating temperature of the early Universe, $T_{\rm PQ}<T_{\rm RH}$, then the strongly charged $Q$ particles will initially be in thermal equilibrium with the SM plasma. After PQ breaking, the heavy quarks acquire a mass and will freeze-out. As pointed out in Refs.~\cite{DiLuzio:2016sbl,DiLuzio:2017pfr}, performing standard thermal production would generate a large relic abundance of heavy quarks, violating constraints from cosmology unless they decay quickly enough. Whether or not $Q$-decay is fast is determined by their SM and PQ charges. For example, the original KSVZ model has a charge configuration such that the heavy quark is stable~\cite{Kim:1979if}. Finding KSZV-type models that enable these heavy quarks to decay fast enough then becomes a model building exercise. \\ 

\noindent Furthermore, the misalignment mechanism generates a cold axion abundance~\cite{Preskill:1982cy, Dine:1982ah, Abbott:1982af, DiLuzio:2020wdo}:
\begin{align}
    \Omega_a h^2\approx 0.12\, \left(\frac{\theta_{\rm i}}{2.15}\right)^{2}\left(\frac{28~\mu {\rm eV}}{m_a}\right)^{7/6}, 
    \label{eq:relic_std}
\end{align}
where $\theta_i$ is the initial angle of the field, $\theta=a/f_a$, which for the $U(1)_{\rm PQ}$ breaking after inflation is approximated to be $\theta_i\approx 2.15$. This number is obtained by accounting for multiple Hubble patches undergoing misalignment with an ensemble of $\theta_i$ values~\cite{GrillidiCortona:2015jxo}. This then limits the axion mass to be $m_a\,\gtrsim 15\, \mu{\rm eV}$ in order to avoid axion overproduction. 

Combining this with the need for fast $Q$ decays, authors of Refs.~\cite{DiLuzio:2016sbl,DiLuzio:2017pfr} utilized an effective field theory approach as a guide on what axion models are compatible with standard cosmology, heavy quark decay terms are described by  
\begin{equation}
    \mathcal{L}_{Qq} = \mathcal{L}_{Qq}^{d\leq 4}+ \mathcal{L}_{Qq}^{d>4}  = \mathcal{L}_{Qq}^{d\leq 4} +\frac{1}{\Lambda^{(d-4)}}\mathcal{O}^{d>4} + {\rm H.c.}\, .
    \label{eq:LEFF}
\end{equation}
where $d$ is the mass dimension of the operators. Taking the cut-off scale as the Planck scale $\Lambda=m_{\rm Pl}=1.22\times10^{19}\,{\rm GeV}$, it was found that the maximum dimension for decay while avoiding the overproduction of axions was $d=5$. This would then exclude models with $Q$ charges with $d\geq 6$ decays at lowest order, such as the model where $Q$ has the SM charges $\left(SU(3)_{\rm c}, SU(2)_{\rm W}, U(1)_{Y}\right)=(3, 1, -7/3)$, which at lowest order has the decay operator $\mathcal{O}=\bar{Q}_L d_R\left(\bar{e}_R^c\,e_R\right)$.  However, recently Ref.~\cite{Cheek:2023fht} found that for decays $d\geq5$, the thermally produced heavy quarks will provoke a period of heavy quark domination (HQD), altering the misalignment mechanism in such a way as to allow for smaller axion masses and larger dimension for $Q$ decay. A period of early matter domination altering the dynamics of the misalignment mechanism has been explored before~\cite{Steinhardt:1983ia, Lazarides:1990xp, Kawasaki:1995vt,Visinelli:2009kt,Nelson:2018via,Ramberg:2019dgi, Arias:2021rer, Xu:2023lxw, Arias:2022qjt, Arias:2023wyg, Mazde:2022sdx}, typically by introducing additional matter content as new physics. Ref.~\cite{Cheek:2023fht} on the other hand pointed out that for KSVZ axion models, the additional matter content is already present. This result means that a greater number of axion models are viable, in Ref.~\cite{DiLuzio:2024xnt} authors enumerate such models. Notably, they found that there are four $d=6$ and two $d=9$ SM charge assignment for the heavy quark that do not suffer from a domain wall problem, this extends the previously known two, KSVZ-I and KSVZ-II, to eight.  

\subsection{Condition for heavy quark domination}
\label{sec:conditions_EMD}

\noindent Here we describe in more detail the thermal freeze-out and later early matter domination of the heavy quarks. The $\bar{Q}+Q\to {\rm SM}+{\rm SM}$ annihilation cross-section is given by 
\begin{equation}
    \langle\sigma v\rangle_{\bar{Q}Q}=\frac{\pi\alpha_s^2}{16 m_Q^2}\left(c_f n_f + c_g\right). \label{eq:FO_BE}
\end{equation}
where $n_f$ is the number of quark flavours that $Q$ can annihilate into, and $\left(c_f, c_g\right)=\left(2/9,\,220/27\right)$ for triplets and $\left(c_f, c_g\right)=\left(2/3,\,27/4\right)$ for octets, in this paper we assume the former. Through the standard thermal freeze-out procedure one can determine the temperature of the SM plasma when $\rho_Q\approx\rho_{\rm SM}$~\cite{Cheek:2023fht}, 
\begin{equation}
    \tdom \approx 10^7 \left( \frac{m_Q}{10^{12} \, \text{GeV}} \right)^2 \, \text{GeV}.
    \label{eq:Teq}
\end{equation}
In order for heavy quarks to dominate the early Universe, their decay must occur after heavy quark-radiation equality. Since the plasma temperature is always decreasing, we require $\tdec<\tdom$. To estimate $T^Q_{\rm dec}$, we set $\Gamma_Q \equiv H_{\rm rad}\left(T=T^Q_{\rm dec}\right)$ where $H_{\rm rad}$ is the Hubble parameter in the radiation-dominated Universe\footnote{ Notice that we have assumed $H_{\rm rad}$ for this approximation, which strictly speaking, is not appropriate during early matter domination. However, when the matter domination is caused by the decaying particle, this estimate holds up-to $\mathcal{O}(1)$ factors. In the later analysis of this article we solve the coupled Boltzmann equations, which show broad consistency with this estimate.},
\begin{equation}
    \tdec^{d}\approx\left(\Gamma^d_Q\right)^{\frac{1}{2}} \left(\frac{1}{g_\star(\tdec)}\right)^{\frac{1}{4}}  \left(\frac{90 \, m_{\rm Pl}^2}{8 \pi^3}\right)^{\frac{1}{4}}. \label{eq:Tdec}
\end{equation}
Where $g_\star$ is the effective energy degrees of freedom. The decay widths of $Q$ for each dimension of decay is 
\begin{align}
    \Gamma^{d=4}_Q &= \frac{m_Q}{8\pi}, \,\,\,\, &\Gamma_Q^{d=5}&= \frac{m_Q^3}{16\pi \Lambda^2},\nonumber\\ 
    \Gamma_Q^{d=6}&= \frac{m_Q^5}{512\pi^3 \Lambda^4},\,\,\,\,  &\Gamma_Q^{d=7}&= \frac{m_Q^7}{49152\pi^5 \Lambda^6}, \,
    \label{eq:EFFwidths}
\end{align}
where we neglect the details of higher dimension decays in this paper. Furthermore, the main focus in this article will be the dimension 6 and 7 dimensional decay terms. This is because $d\le5$ models will not produce periods of HQD that will alter the misalignment mechanism. Models with $d=5$ decays can lead to HQD, but these typically require $m_Q>f_a$ or predict overproduction of the axion due to a large $f_a$ and has no in-built mechanism to dilute it. According to Ref.~\cite{DiLuzio:2024xnt}, there are 43 models that have dimension 6 decays at lowest order and 44 for dimension 7. Using Eq.~(\ref{eq:EFFwidths}) one can estimate when $\tdom>\tdec$. For $d=6$ and $d=7$, this is when
\begin{eqnarray}
    \Lambda &\gtrsim& (4\times10^{28} \,{\rm GeV}^2\,\times m_Q M_{\rm Pl})^{1/4}\,\,\,\,{\rm and}\nonumber \\
    \Lambda &\gtrsim& (4\times10^{25} \,{\rm GeV}^2\,\times m_Q^3 M_{\rm Pl})^{1/6},
\end{eqnarray}
respectively. Furthermore, with the calculation of $\tdec$ we are able to assess whether a particular model and parameter point is in tension with BBN~\cite{Kawasaki:2004qu, Jedamzik:2006xz, Jedamzik:2007qk, Kawasaki:2017bqm}. Typically one estimates this by checking that $\tdec\gtrsim 3\,{\rm MeV}$.\\

\subsection{Axion misalignment with heavy quark domination}
\label{subsec:axion_cosmo}

The details of the axion misalignment mechanism within non-standard cosmologies can be found in Refs.~\cite{Steinhardt:1983ia, Lazarides:1990xp, Kawasaki:1995vt,Visinelli:2009kt,Nelson:2018via,Ramberg:2019dgi, Arias:2021rer, Xu:2023lxw, Chen:2025awt}. The insight of Ref.~\cite{Cheek:2023fht} was that axion models themselves typically contain all necessary ingredients for early matter domination.

The full dynamics are determined by the coupled Friedmann Boltzmann equations 
\begin{subequations}\label{eq:FBEqs}
    \begin{eqnarray}
    \frac{3H^2M_{\rm Pl}^2}{8\pi}&=&\rho_\mathrm{R}^{\rm SM} + \rho_{Q}\,,\\
     \frac{\dd s_\mathrm{R}^{\rm SM}}{\dd t} &=& -3H s_\mathrm{R}^{\rm SM}  +\frac{{\rm BR}_{\rm SM}\Gamma_Q}{T}\rho_{Q} \,,\\
     \frac{\dd n_{Q}}{\dd t} &=& - 3H n_{Q} -\Gamma_Q n_Q -\langle\sigma\,v\rangle\left[n_Q^2 -(n_Q^{\rm eq})^2\right] \,,\nonumber\\
    \end{eqnarray}
    \label{eq:FBEqs_FEq}
    \end{subequations}
\noindent where $\rho$, $n$ and $s$ are the energy, number, and entropy densities respectively, we also denote each component, SM radiation ($\textrm{R}$), heavy quark ($Q$), in the subscript.  
Note that we haven't included the Boltzmann equation for the the thermally produced axion. As shown in Refs.~\cite{Cheek:2023fht, Cheek:2024ofn}, if HQD occurs, the branching ratio to axions must be subdominant otherwise axion contributions to $\Delta N_{\rm eff}$ would be too large. Hence, in this work we only consider 100\% decay into SM particles, which implies that if HQD occurs and ends after thermal axion decoupling, then $\Delta N_{\rm eff}$ will be much less than the expected value from the standard thermal axion, \textit{i.e.} $\Delta N_{\rm eff}\ll 0.027$. This is despite the axion actually having been in thermal contact with the SM early on.

One can solve Eqs.~(\ref{eq:FBEqs}) to validate the approximations made in Eq.~(\ref{eq:Teq}) and Eq.~(\ref{eq:Tdec}), and to when generate the GW signals in section~\ref{sec:pgw}. We use the precise determination of $H$ as an input to solve the misalignment mechanism using the dedicated solver {\tt MiMeS}~\cite{Karamitros:2021nxi}. Specifically, we solve the equation of motion for the axion field, $a$, which is expressed in terms of the axion angle, $\theta= a/f_a$
\begin{equation}\label{eq:misalignment}
    \left(\frac{\dd^2}{\dd t^2} + 3H(t)\frac{\dd}{\dd t}\right)\theta(t) + \tilde{m}_a^2(t)\sin(\theta(t))=0\, , 
\end{equation}
where $\tilde{m}_a(t)$ is the thermal mass of the axion~\cite{Borsanyi:2016ksw}. Generically, the axion abundance is diluted such that a greater $f_a$ is required to achieve the correct relic abundance for axion dark matter. 

In order to make the connection with the experimental search for axion dark matter, we numerically solve Eqs.~(\ref{eq:FBEqs}) and Eq.~(\ref{eq:misalignment}) to find the required $f_a$ such that $\Omega_a\,h^2=\Omega_{\rm DM}\,h^2$, where we take the relic abundance of dark matter determined in~\cite{Planck:2018vyg}. We do this for a range of $m_Q$ and $\Lambda$ values, and take $\theta_i=2.15$ in-line with calculations of post-inflationary PQ breaking~\cite{GrillidiCortona:2015jxo}. Furthermore, we check that $\rho_{\rm R}>\rho_{Q}$ at the onset of BBN. In previous works $m_Q\sim f_a$ and $\Lambda \sim m_{\rm Pl}$ were taken, but these are simply the upper bounds on such parameters, we find that by insisting that the QCD axion has to account for all of the dark matter, these assumptions can be relaxed and we still have a sufficiently constrained scenario. 

\begin{figure}[!ht]
    \begin{center}
        \includegraphics[width=0.9\linewidth]{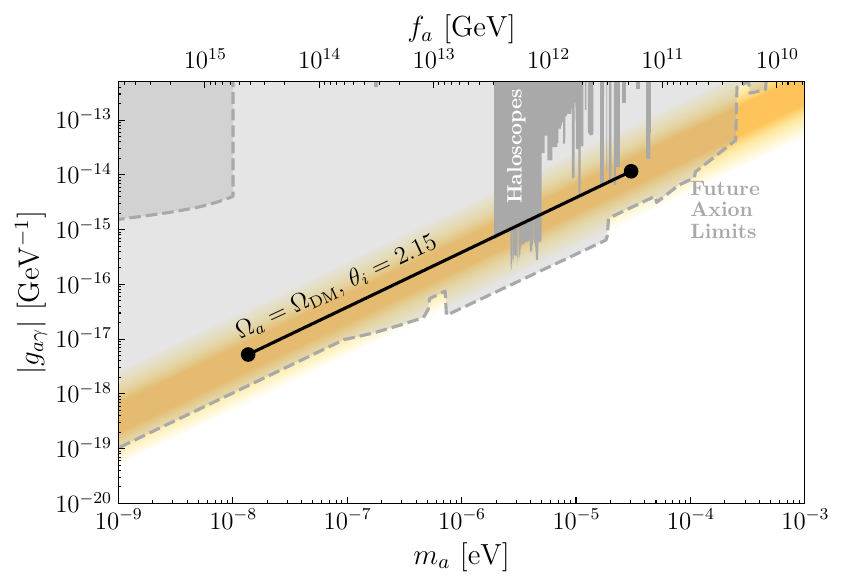}
        \caption[]{\justifying \it The range of axion mass $m_a$ which can reproduce the correct relic abundance allowing for any $m_Q\lesssim f_a$ and $\Lambda<m_{\rm Pl}$. This is for $d=6$ but we find the result for $d=7$ decays to be very similar. The axion limits and QCD axion band was plotted using Ref.~\cite{AxionLimits}.}
        \label{fig:EMD_d6_d7}
    \end{center}
\end{figure}

Fig.~\ref{fig:EMD_d6_d7} shows the range of $m_a$ values that can achieve the correct relic abundance, allowing $m_Q$ and $\Lambda$ to vary. We have presented the possible $m_a$ range in the $\left(m_a, g_{a\gamma}\right)$-plane, where $g_{a\gamma}$ is the axion-photon coupling which depends on the anomaly coefficients, $N$, from Eq.~(\ref{eq:LCP}) and $E$, the coefficient of the electromagnetic anomaly term, 
\begin{equation}
    g_{a\gamma}\equiv \frac{\alpha}{2\pi}\frac{1}{f_a}\left(\frac{E}{N}-1.92(4)\right).
\end{equation}
$\alpha$ is the electromagnetic fine-structure constant. The precise value for $E/N$ depends on the SM and PQ charges of the HQs, in other words the specific axion model one is considering. To encompass the range of $E/N$ values we show a yellow band. To represent the range of $m_a$ where the correct relic abundance is achieved we chose the original KSVZ model, $E/N=0$, which is the black line connecting the two black points. However, we emphasize that axion models of interest are in the range of $E/N$ values that lead to the range $\left\vert E/N-1.92(4)\right\vert \in\left[0.003-100\right]$. We note that by allowing $m_Q\lesssim f_a$ we have found a substantial quantitative difference from the results in Refs.~\cite{Cheek:2023fht,DiLuzio:2024xnt} where models with $d=6$ decay terms were found to only marginally change the expected $m_a$ value that reproduces the correct relic density, even when $\Lambda$ could vary. In Fig.~\ref{fig:EMD_d6_d7} we see that $d=6$ decays can alter the cosmology such that the $f_a$ required for $\Omega_{a}\, h^2=\Omega_{\rm DM}\,h^2$ is orders of magnitude higher. What limits this value is the constraints from BBN. The left most point on the axion dark matter line corresponds to $f_a=4\times10^{14}\,{\rm GeV}$, $m_Q=4\times10^{11}\,{\rm GeV}$, and $\Lambda=4\times10^{18}\,{\rm GeV}$. The right most point corresponds to the expectation for a dark matter axion without HQD. Calculating $T_{\rm dec}$ via Eq.~(\ref{eq:Tdec}) gives $5\,{\rm MeV}$, but due the more precise numerical evaluation from Eq.~(\ref{eq:FBEqs_FEq}) gives $T_{\rm dec}=3\,{\rm MeV}$. 

\begin{figure*}
    \centering
    \includegraphics[width=1.0\linewidth]{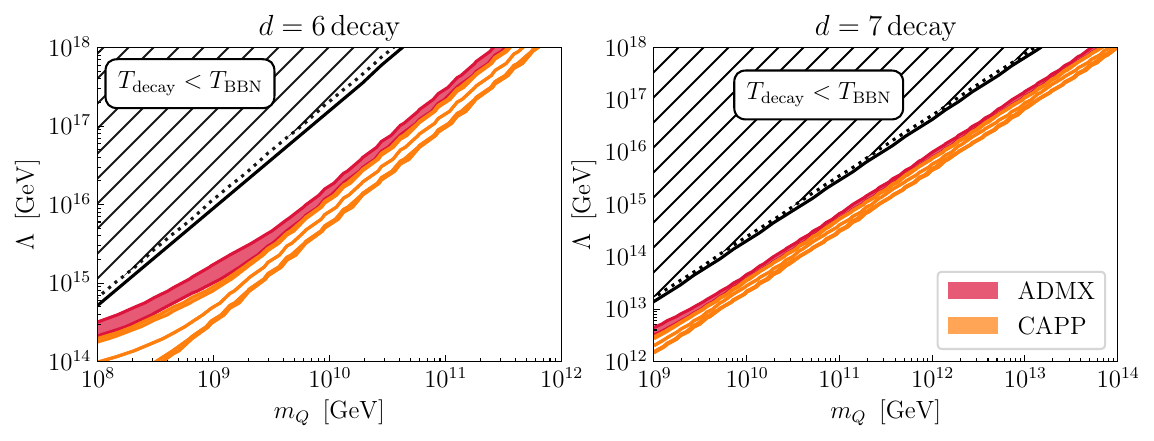}
    \caption[]{\justifying \it The current constraints on the $m_Q$ and $\Lambda$ parameter space from ADMX and CAPP experiments. The left panel is for $d=6$ decays and the right panel is for $d=7$ decays. The value of $f_a$ is not a priori set, instead we determine the required value such that $\Omega_a=\Omega_{\rm cdm}$. The hatched region is where $Q$ decay occurs too late and will be constrained by BBN, we have bounded it by the dotted and solid black lines which correspond to the more approximate Eq.~(\ref{eq:Tdec}) and precise numerical evaluation via Eqs.~(\ref{eq:FBEqs_FEq}) respectively.}
    \label{fig:current_experiments}
\end{figure*}

We show this slight difference in the way we calculate the constraint from BBN on the $\left(m_Q, \Lambda\right)$-plane in Fig.~\ref{fig:current_experiments}. The black and white hatched region is bounded by a dotted and a solid line, which indicates the approximate bound (Eq.~(\ref{eq:Tdec})) or full numerical one (Eqs.~(\ref{eq:FBEqs_FEq})) respectively. One can see the constraints coming from the more accurate numerical evaluation is more aggressive, henceforth the BBN bounds we will show correspond to this. Furthermore, a more accurate determination is possible by combining our Boltzmann equations with those of the light elements~\cite{Kawasaki:2017bqm}, we leave this for future work. All the points in Fig.~\ref{fig:current_experiments} to the right of the hatched region can satisfy $\Omega_{a}=\Omega_{\rm DM}$ given the certain choice of $f_a$. Some of those $f_a$ values will coincide with the haloscope experiment constraints shown in Fig.~\ref{fig:current_experiments} as a dark gray region.  In particular, the two experiments that have reached sensitivities down to the KSVZ line in the relevant $f_a$ region is ADMX~\cite{ADMX:2009iij,Stern:2016bbw} and CAPP~\cite{CAPP:2020utb, Yang:2023yry, Kim:2023vpo, CAPP:2024dtx}, they are the crimson red and orange in shaded regions in Fig.~\ref{fig:current_experiments}. Note that since these experiments are haloscope experiments, their sensitivity is intimately tied to the $\Omega_a=\Omega_{\rm DM}$ requirement we made. In the left panel of Fig.~\ref{fig:current_experiments} we show the results for $d=6$ decay terms, and in the right panel we show the results for $d=7$ decay terms. This will be useful in assessing the potential for future GW experiments to probe the same parameter space. As we will show in section~\ref{sec:pgw}, the inflationary GW signal is sensitive $m_Q$ and $\Lambda$ and not $f_a$ directly.

Before we move on, we comment that we have not included any contribution to the axion abundance from topological defects such as cosmic strings or unstable domain walls~\cite{Kibble:1976sj,Kibble:1982dd,Vilenkin:1982ks}. We do this because the exact contribution is still under dispute in the literature and all calculations to date have been performed under standard cosmology~\cite{Hagmann:2000ja,Wantz:2009it, Hiramatsu:2010yu, Kawasaki:2014sqa, Buschmann:2021sdq, Gorghetto:2021fsn,Kim:2024wku}. Furthermore, since the vast majority of the axion abundance is produced around the QCD phase transition $T_{\rm QCD}$, similarly to the misalignment mechanism, the effect where HQD dilutes the axion abundance will also hold. Future work is warranted in this direction, but it is beyond the scope of this article.

\subsection{Interpreting the cosmology of a low mass axion} 

In the previous subsection we have shown that axion models with post-inflationary PQ breaking can lead to axion dark matter with masses as low as $\sim 10^{-8}\,{\rm eV}$. The only requirements are that the new particles that generate the anomaly term in Eq.~(\ref{eq:LagPQ}), are heavy ($m_Q\gtrsim10^7\,{\rm GeV}$) and live long enough. For the effective parametrisation of Eq.~(\ref{eq:LEFF}), this translates to the requirement that the lowest dimension decay terms have to be $d\geq 6$. What's more, recently Ref.~\cite{DiLuzio:2024xnt} found that there exists four dimension 6 KSVZ-type charge configurations that do not have a domain wall problem, for lower dimensions, there are only two configurations. Additionally, it appears that these new models contain $Q$ decays to SM particles only, therefore $\Delta N_{\rm eff}\ll 0.027$~\cite{Cheek:2024ofn}. 

This poses an interesting problem for distinguishing between the pre- and post-inflationary PQ breaking scenarios because now both possibilities can easily produce the same values for $m_a$ and $\Delta N_{\rm eff}$. What's more, the pre-inflationary $U(1)_{\rm PQ}$ breaking case typically requires a finely tuned initial angle $\theta_i\ll1$ to produce axion dark matter with masses $m_a\lesssim10^{-6}\,{\rm eV}$. In the context of post-inflationary breaking we presented above, fine-tuning comes in via small Yukawa couplings, $y_q\sim10^{-3}$, well within the range observed in the SM, so shouldn't cause much theoretical consternation. 

As always, we would like to find experimental avenues to break the degeneracy between the pre and post-inflationary breaking scenario. What's more, some kind of understanding of when $U(1)_{\rm PQ}$ breaking occurred relative to the reheating temperature would constitute a profound insight into the nature and scale of inflation. As we've seen in Eq.~(\ref{eq:Teq}) and Eq.~(\ref{eq:EFFwidths}) the onset and end of HQD is sensitive to fundamental parameters of the underlying theory, so if we can probe these events experimentally in some way, we will be able to learn about the possible axion models. As we will see in the next section, inflationary gravitational waves may provide precisely this insight.

\section{Primordial Gravitational Waves}
\label{sec:pgw}

An early matter domination epoch, occurring within post-reheating and pre-BBN era leaves distinct imprints on stochastic gravitational wave background. Here we will explore how the HQD influences the GW spectrum, examining the unique features it induces. GWs generated during inflation exhibit constant amplitudes outside the horizon and undergo damping upon re-entry. At the present epoch, the observed GW spectrum, $\Omega_\text{GW}(k)$, can be expressed in terms of wave-number ($k$) as
\begin{align}
    \label{eq:omega}
    \Omega_\text{GW}(k) = \frac{1}{12}\left(\frac{k}{a_0 H_0}\right)^2 T_T^2(k) \;P_T^\text{prim}(k),
\end{align}
with $a_0 (H_0)$ being the Friedmann-Lema\^itre-Robertson-Walker (FLRW) metric scale factor (Hubble parameter) at the present time~\cite{Datta:2022tab}\footnote{Throughout our study $a_0=1$ and $H_0=67$ km/s/Mpc.}. $P_T^\text{prim}$ represents the primordial power spectrum from inflation which can be expressed as~\cite{Meerburg:2015zua,Liddle:1993ez,LIGOScientific:2016jlg,  LIGOScientific:2006zmq,Cabass:2015jwe,Allen:1997ad,LIGOScientific:2011yag}
\begin{align}
    \label{eq:pri_tensor}
    P_T^\text{prim.}(k) = A_T(k) \left(\frac{k}{k_{\ast}}\right)^{\nt}.
\end{align}
$A_T$ represents the amplitude of the tensor power spectra and determines the amplitude of the GW spectrum. It is evaluated at a particular pivot scale $k_{\ast}=0.05$ Mpc$^{-1}$, where the Planck collaboration reports its observational constraints~\cite{Planck:2018vyg}.
Additionally, $\nt$ is the tensor spectral index, which governs the tilts of the GW spectrum. A positive $\nt$ corresponds to a blue-tilted spectrum, while a negative $\nt$ leads to red-tilted spectrum. Although standard single-field slow-roll inflation generally predicts a red-tilted $\nt$~\cite{Liddle:1993fq}, several alternative scenarios can produce a blue-tilted $\nt$ spectrum, for instances, string cosmology \cite{Brandenberger:2006xi,Calcagni:2013lya}, modified gravity~\cite{Fujita:2018ehq}, particle production during inflation~\cite{Cook:2011hg,Mukohyama:2014gba}, G-inflation~\cite{Kobayashi:2010cm}, Higgsed chromo-natural inflation~\cite{Adshead:2016omu}, natural inflation on steep potentials~\cite{Anber:2009ua}, super-inflation modes~\cite{Baldi:2005gk}, Einstein-Gauss-Bonnet inflation~\cite{Oikonomou:2021kql}, spectator scalar field during the inflation~\cite{Biagetti:2013kwa}, Axion gauge field inflation~\cite{Caldwell:2017chz,Dimastrogiovanni:2018xnn}, Higgsed gauge-flation~\cite{Adshead:2017hnc}. Furthermore, recent NANOGrav results have been interpreted as a blue-tilted spectrum, showing a good fit~\cite{Kuroyanagi:2020sfw,NANOGrav:2023hvm}.

Blue-tilted spectra can overproduce gravitons, that will contribute to  the additional effective number of relativistic degrees of freedom, $\Delta N_{\rm eff}$~\cite{Maggiore:1999vm}
\begin{eqnarray}\label{eq:omega_delta_nu}
    \int^{\infty}_{f_{\rm min}} \frac{df}{f} \Omega_{\rm GW}(f)h^2\;\leq 5.6\times 10^{-6}\;\Delta N_{\rm eff}, 
\end{eqnarray}
where $f_{\rm min}$ is the lower limit of integration, which is $\simeq \; 10^{-10}\, (\simeq \; 10^{-18})$ Hz for BBN (CMB). Current observational constraints from Planck 2018 and BAO~\cite{Planck:2018vyg}, along with BBN~\cite{Cyburt:2015mya}, place stringent upper bounds on $\Delta N_{\rm eff}$. Future CMB experiments, including CMB Bharat~\cite{CMB-bharat}, CMB-S4~\cite{Abazajian:2019eic,TopicalConvenersKNAbazajianJECarlstromATLee:2013bxd}, PICO~\cite{NASAPICO:2019thw}, CMB-HD~\cite{CMB-HD:2022bsz}, COrE~\cite{CORE:2017oje}, the South Pole Telescope~\cite{SPT-3G:2014dbx}, and the Simons Observatory~\cite{SimonsObservatory:2018koc}, are expected to further refine these constraints with improved sensitivity. Among current observations, Planck 2018+BAO provides one of the most stringent upper bounds on $\Delta N_{\rm eff}$, limiting it to $0.28$~\cite{Cyburt:2015mya}. Looking ahead, CMB-HD is projected to significantly improve this constraint, reaching a sensitivity of $0.027$~\cite{CMB-HD:2022bsz}. In this study, we incorporate both the current Planck 2018+BAO limit and the projected CMB-HD constraint to analyse their implications.

$T_T(k)$ in Eq.~\eqref{eq:omega} is the transfer function describing the evolution of gravitational waves throughout different epochs, in the background of a FLRW Universe. In general, it can be expressed as~\cite{Turner:1993vb,Chongchitnan:2006pe,Nakayama:2008wy,Nakayama:2009ce,Kuroyanagi:2011fy,Kuroyanagi:2014nba}:
\begin{align}
\label{eq:transfer}
    T_T^2(k) \equiv \Omega_m^2 \left(\frac{g_*(T_\text{in})}{g_*^0}\right)\left(\frac{g_{*S}^0}{g_{*S}(T_\text{in})}\right)^\frac{4}{3} \left(\frac{3j_1(z_k)}{z_k}\right)^2 F(k),
\end{align}
where $g_{\star\,S}$ is the total number of entropy degrees of freedom and $\Omega_m (=0.31)$ is the total matter density~\cite{Planck:2018vyg}. $j_1(z_k)$ is the first order spherical Bessel function, with $z_k\equiv k\tau_0$, where $\tau_0=2 / H_0$ \cite{Datta:2022tab,Planck:2018vyg} being the conformal time today. Now, the imprints of any non-standard evolution during the pre-BBN epoch are encapsulated within $F(k)$. In the standard scenario (\textit{i.e.} only radiation domination after inflationary reheating and before BBN), $F(k)$ has the following form:
\begin{align}\label{eq:stand}
    F(k)\bigg |_\text{standard} \equiv T_1^2\left(\frac{k}{\keq}\right)T_2^2\left(\frac{k}{\krh}\right),
\end{align}
where $k_{\rm eq}$ and $k_{\rm RH}$ correspond to modes that reenter horizon at matter-radiation equality and the end of reheating respectively,
\begin{eqnarray}
    \keq &=& 7.1\times 10^{-2} \Omega_m h^2\,{\rm Mpc}^{-1}, \label{eq:keq}\\
    \krh &=& 1.7\times 10^{14}\left(\frac{g_{*S}(\trh)}{g_{*S}^0}\right)^\frac{1}{6} \left(\frac{\trh}{10^7\;\text{GeV}}\right)\,{\rm Mpc}^{-1}.\nonumber\\ \label{eq:krh}
\end{eqnarray}
In the case of HQD, $F(k)$ becomes
\begin{eqnarray}
\label{eq:IMD}
    F(k)\bigg |_\text{HQD} \equiv &  
      T_1^2\left(\frac{k}{\keq}\right)T_2^2\left(\frac{k}{\kdec}\right)\times\nonumber \\
      &T_3^2\left(\frac{k}{\kdecs}\right)T_2^2\left(\frac{k}{\krhs}\right),
\end{eqnarray}
where $\kdec$ corresponds to the wave-number at the end of HQD, estimated as 
\begin{eqnarray}
\label{eq:kdec}
    \kdec &= 1.7\times 10^{14} \left(\frac{g_{*S}(\tdec)}{g_{*S}^0}\right)^\frac{1}{6} \left(\frac{\tdec}{10^7\;\text{GeV}}\right)\, {\rm Mpc}^{-1}.
\end{eqnarray}
The characteristic wave-numbers $\kdecs$ and $\krhs$, in Eq.~\eqref{eq:IMD} can be expressed as
\begin{eqnarray}
    \kdecs &\equiv& \kdec \Delta_s^{2/3},\label{eq:kdecs}\\
    \krhs &\equiv& \krh \Delta_s^{-1/3}\label{eq:krhs},
\end{eqnarray}
where $\Delta_s$ is the entropy dilution factor which quantifies the the change of comoving entropy density due to the decay of the heavy quark. It is defined as the ratio of comoving entropy density after HQD to that before its onset, \textit{i.e.}, $\Delta_s\equiv S(\tdec)/S(\tdom)$. To accurately determine this dilution factor, we numerically solved the coupled Friedmann Boltzmann equations, given in Eqs.~\eqref{eq:FBEqs}, ensuring a precise treatment of the entropy evolution throughout the process. 

The fitting functions ($T_1^2(x), T_2^2(x)$ and $T_3^2(x)$) in Eq.~\eqref{eq:transfer} can be expressed as~\cite{Kuroyanagi:2014nba} 
\begin{eqnarray}
    T_1^2(x) &\equiv& 1+1.57 x +3.42 x^2,\label{eq:T1}\\
    T_2^2(x) &\equiv& (1-0.22x^{3/2}+0.65x^2)^{-1},\label{eq:T2}\\
    T_3^2(x) &\equiv& 1+0.59 x +0.65 x^2\label{eq:T3}.
\end{eqnarray}
Hence, using the above set of equations we can estimate the nature of GW spectra for the different axion parameters, leading to the HQD epoch.

\subsection{GW Detection prospects with interferometers}
\label{subsec:detection_pros}

The inflationary GW signals described in the previous subsection are yet to be detected. With the exceptional progress in GW interferometry, the prospects for observing them have significantly improved. The experimental landscape we consider can be categorized in the following way:
\begin{enumerate}[a)]
    \item \textbf{Terrestrial detectors:} \textit{Laser Interferomenter Gravitational-wave Observatory} (LIGO)~\cite{LIGOScientific:2016aoc,LIGOScientific:2016sjg,LIGOScientific:2017bnn,LIGOScientific:2017vox}, \textit{Einstein Telescope} (ET)~\cite{Punturo_2010,Hild:2010id}, \textit{Advanced} LIGO (a-LIGO)~\cite{LIGOScientific:2014pky,LIGOScientific:2019lzm}, LIGO-India~\cite{Iyer2011,article,Unnikrishnan:2013qwa,Sathyaprakash2012,Unnikrishnan:2023uou}, VIRGO~\cite{LIGOScientific:2010ped,2012JInst...7.3012A,LIGOScientific:2017ycc,LIGOScientific:2017vwq}, advanced-VIRGO ~\cite{PhysRevLett.123.231108,VIRGO:2014yos}~\cite{KAGRA:2018plz}, \textit{Cosmic Explorer} (CE)~\cite{Reitze:2019iox}
    \item \textbf{Space based detectors: }\textit{Deci-Hertz Interferometer Gravitational-wave Observatory} (DECIGO)~\cite{Yagi:2011yu}, \textit{Upgraded} DECIGO (U-DECIGO)~\cite{Seto:2001qf,Kawamura_2006,Yagi:2011wg}, \textit{Big-Bang Observer} (BBO)~\cite{Corbin:2005ny,Harry_2006},  \textit{Laser Interferometer Space Antenna} (LISA)~\cite{amaroseoane2017laser,Baker:2019nia}, $\mu$-ARES~\cite{Sesana:2019vho}. 
    \item \textbf{Pulsar timing arrays:} \textit{Square Kilometre Array} (SKA)~\cite{Carilli:2004nx,Janssen:2014dka,Weltman:2018zrl}, \textit{European Pulsar Timing Array} (EPTA)~\cite{Kramer:2013kea,Lentati:2015qwp,Babak:2015lua}, \textit{North American Nanohertz Observatory for Gravitational Waves} (NANOGrav)~\cite{McLaughlin:2013ira,NANOGRAV:2018hou,Aggarwal:2018mgp,Brazier:2019mmu,NANOGrav:2020bcs}
    \item \textbf{Cosmic microwave background radiation:} Detecting BB-modes in CMB with current and future satellites. Current observational constraints from Planck 2018~\cite{Planck:2018vyg}. Future CMB experiments include CMB Bharat~\cite{CMB-bharat}, CMB-S4~\cite{Abazajian:2019eic,TopicalConvenersKNAbazajianJECarlstromATLee:2013bxd}, PICO~\cite{NASAPICO:2019thw}, CMB-HD~\cite{CMB-HD:2022bsz}, COrE~\cite{CORE:2017oje}, the South Pole Telescope~\cite{SPT-3G:2014dbx}, and the Simons Observatory~\cite{SimonsObservatory:2018koc}.
\end{enumerate} 

\begin{figure*}[!ht]
    \centering
    \includegraphics[scale=0.45]{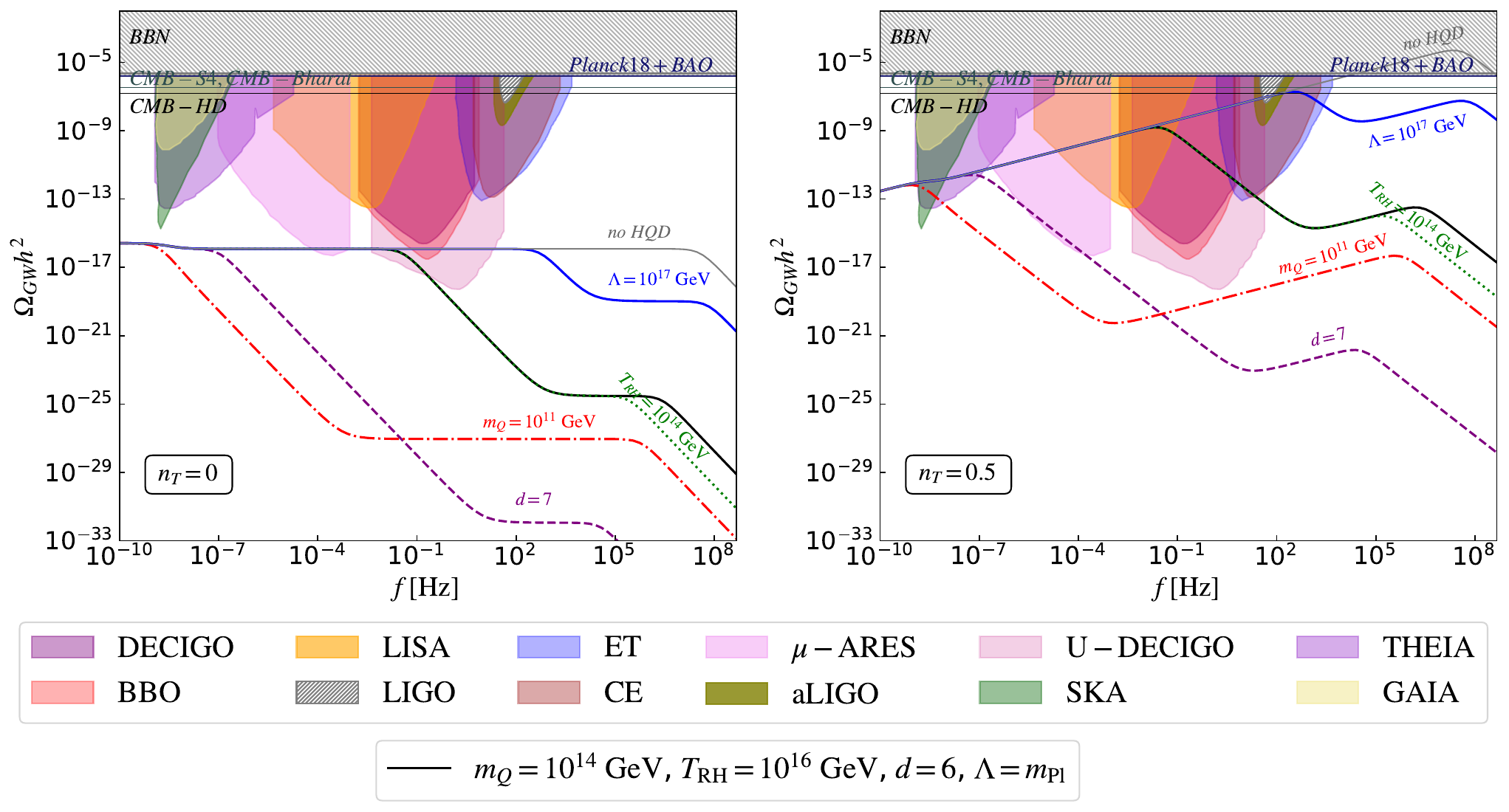}
    \caption[]{\justifying\it Illustration of GW spectrum for different values of $m_Q$, $d$ and $\trh$, compared with the standard scenario (without HQD) via thin grey solid line. In each of the GW spectrum (dashed, dot-dashed, dotted), variation with respect to parameters are quoted. Grey hatched regions are the excluded regions by the existing constraints (Planck-18+BAO, BBN and LIGO). \textbf{Left panel} is for $\nt = 0$, whereas \textbf{Right panel} is for $\nt=0.5$. Here $m_Q$ and $\trh$ are in GeV. For both of the plots $r=0.036$ is fixed which corresponds to the Hubble scale of inflation $H_{\rm inf}\sim10^{13}$ GeV. }
    \label{fig:GW_spectrum}
\end{figure*}
In Fig.~\ref{fig:GW_spectrum}, we illustrate the behaviour the GW spectrum in presence of the HQD, as a function of frequency, assuming $r=0.036$~\footnote{$r=0.036$ represents the present upper bound on $r$ from Planck-18 + BK18 data, measured at the pivot scale~\cite{BICEP:2021xfz,BICEP2:2015nss}.}. 
The left panel of Fig.~\ref{fig:GW_spectrum} corresponds to $\nt=0$, where the prospects of detecting the signal with current and future detectors remains low. However a blue-tilted ($\nt>0$) spectrum is able to enhance the detection prospects, as depicted in the right panel of Fig.~\ref{fig:GW_spectrum}.

Additionally, in Fig.~\ref{fig:GW_spectrum}, we show the power-law integrated (PLI) sensitivity curves~\cite{Thrane:2013oya} for various GW experiments, gray shaded regions are the existing bounds while rest are future sensitivity projections. These curves assume a GW spectrum varies with frequency as $\Omega_{\rm GW}\sim f^b$, with $b$ being the power-law exponent. The region enclosed by the PLI curve typically indicates the parameter space where a power-law signal can be detected with robust signal-to-noise ratio (SNR). Although this provides a useful benchmark, robust detection is possible only for PLI integrated curves. Therefore, we rely on a direct SNR computation to assess detectability for specific model parameters.

The PLI sensitivity curves regarding gravitational wave observatories are shown in Fig.~\ref{fig:GW_spectrum}.
In particular, we show ground-based ET~\cite{Punturo_2010,Hild:2010id}, space-based LISA~\cite{amaroseoane2017laser,Baker:2019nia} interferometer GW missions as well as the next generation space-based experiments like BBO~\cite{Corbin:2005ny,Harry_2006} and DECIGO~\cite{Yagi:2011yu}. We also show PTAs, including the futuristic ones like the SKA~\cite{Carilli:2004nx,Janssen:2014dka,Weltman:2018zrl} telescope. 

In both panels of Fig.~\ref{fig:GW_spectrum}, we present the behaviour of GW spectra for the four distinct HQD parameters, compared with standard scenario \textit{i.e.} absence of HQD by thin grey solid line. The black solid line represents a benchmark parameter set with $m_Q=10^{14}$ GeV, $d=6$, $\trh=10^{16}$ GeV and $\Lambda=m_{\rm Pl}$. The other four curves (blue dashed, green dotted, red dot-dashed and purple short-long dashed) represent the variation of single parameters, annotated in the figure with corresponding colour, while keeping the others fixed. For instance, red dot-dashed line corresponds to $m_Q=10^{11}$ GeV while $d$, $\trh$ and $\Lambda$ remains unchanged from the black solid line. In absence of HQD, GW spectra has only one suppression which marks the matter-dominated inflationary reheating phase. The presence of HQD phase, GW spectra exhibit an additional suppression at the lower frequencies, marking the period of heavy quark decay, whereas the higher one corresponds to the same, matter-dominated inflationary reheating with $k_{\rm dec}<k_{\rm RH}$. Thus, by suppressing the GW spectrum, the presence of HQD allows higher values of $\nt$ compared to the standard scenario (absence of HQD).
The duration of HQD and the starting point of the suppression depend on both $\tdom$ and $\tdec$ where $\tdom$ depends only on $m_Q$, while $\tdec$ is affected by $m_Q$, $d$ and $\Lambda$.
More quantitively, $f_{\rm dec}=\kdec/(2\pi)$, where $k_{\rm dec}$ is related to $\tdec$ (Eq.~\eqref{eq:kdec}). At higher $k$ the GW spectrum is effected by entropy dilution until $f_{\rm dec,S}=\kdecs/(2\pi)$ (Eq.~\eqref{eq:kdecs}). During this period, as the Universe is dominated by matter, GW spectrum scales as $f^{-2}$. For instance, the black solid line in both figure exhibits a fall around $f=f_{\rm dec}\approx3\times 10^{-5}\,{\rm Hz}$, which persists up to $f=f_{\rm dec,S}\approx 10$ Hz.
The suppression of the GW spectrum at lower frequencies becomes more dominant as the duration of HQD increases which of course depends on $\Gamma_Q^d$. For a particular $m_Q$ and $d$, $\Gamma_Q^d$ decreases as $\Lambda$ increases~(Eq.~\eqref{eq:EFFwidths}) suppressing GW spectra, this can be seen in Fig.~\ref{fig:GW_spectrum}. By increasing $m_Q$ the decay rate increases as does $\tdom$. However, the suppression becomes less for the higher values of $m_Q$ and $d$, as seen in Fig.~\ref{fig:GW_spectrum} and understood from Eq.~\eqref{eq:EFFwidths}.

Here it is crucial to mention that PQ symmetry breaking and associated axion cosmology also leads to the formation of topological defects like global cosmic strings and domain walls which radiate GWs (see Refs.\cite{Gorghetto:2021fsn,Fu:2023nrn}) and may also collapse into Primordial Black Holes \cite{Ferrer:2018uiu,Conaci:2024tlc}. However, they are only relevant when PQ symmetry breaking scale $f_a$ is of the order of $10^{14}$ GeV or higher. This represents the upper bound of the parameter space what we investigate in this article (see Fig.~\ref{fig:EMD_d6_d7_GW}). Moreover given the period of early HQD domination such GW signals will get further suppressed leading to almost negligible signatures compared to GW from inflation that we study here, see Ref.~\cite{Ghoshal:2023sfa,Datta:2025yow} for details. Nevertheless a detailed analysis of such GW analysis combined with IGW may be interesting but is beyond the scope of the present work.

\subsection{Estimation of signal-to-noise ratio (SNR)}
\label{subsec:snr}
In any experimental setup, noise from several grounds affects the measured data which demands a precise characterization of the noise spectra. Thus, to quantify the prospects of signal detection in a given experiment, the SNR serves as a fundamental diagnostic tool. For any GW experiment, the SNR can be defined as~\cite{Thrane:2013oya,Caprini:2015zlo}
\begin{eqnarray} 
\label{eq:SNR}
     {\rm SNR} \equiv \sqrt{\tau_{\rm obs} \int_{f_{\rm min}}^{f_{\rm max}} \text{d}f \left(\frac{ \Omega_{\rm GW}(f,\{\theta\}) h^2}{\Omega_{\rm GW}^{\rm noise}(f) h^2}\right)^2 },  
\end{eqnarray}
where $\Omega_{\rm GW}(f,\{\theta\})$ denotes the theoretical energy density spectrum of GW signal, characteristically parametrized by $\{\theta\}$ and $\tau_{\rm obs}$ represents the observation period of the detectors. The term $\Omega_{\rm GW}^{\rm noise}(f)$ characterizes the detector's noise power spectral density, with $f_{\rm min}$ and $f_{\rm max}$ being the operating frequency range for GW detectors. 

To ensure broad spectral coverage, we consider multiple GW detectors, including SKA, $\mu$-ARES, LISA, BBO, and ET, collectively spanning the frequency range $[10^{-9},10^3]$ Hz. The detailed analysis for noise characteristics of these detectors are discussed in Appendix~\ref{app:noise}. For consistency in our analysis, we have adopted a uniform detection threshold of ${\rm SNR} = 10$ across all GW experiments to evaluate their sensitivity to potential signals. The relevant detector specifications are summarized in Table-\ref{tab:detector_spec}.

\begin{figure*}[!ht]
    \centering
    \includegraphics[scale=0.39]{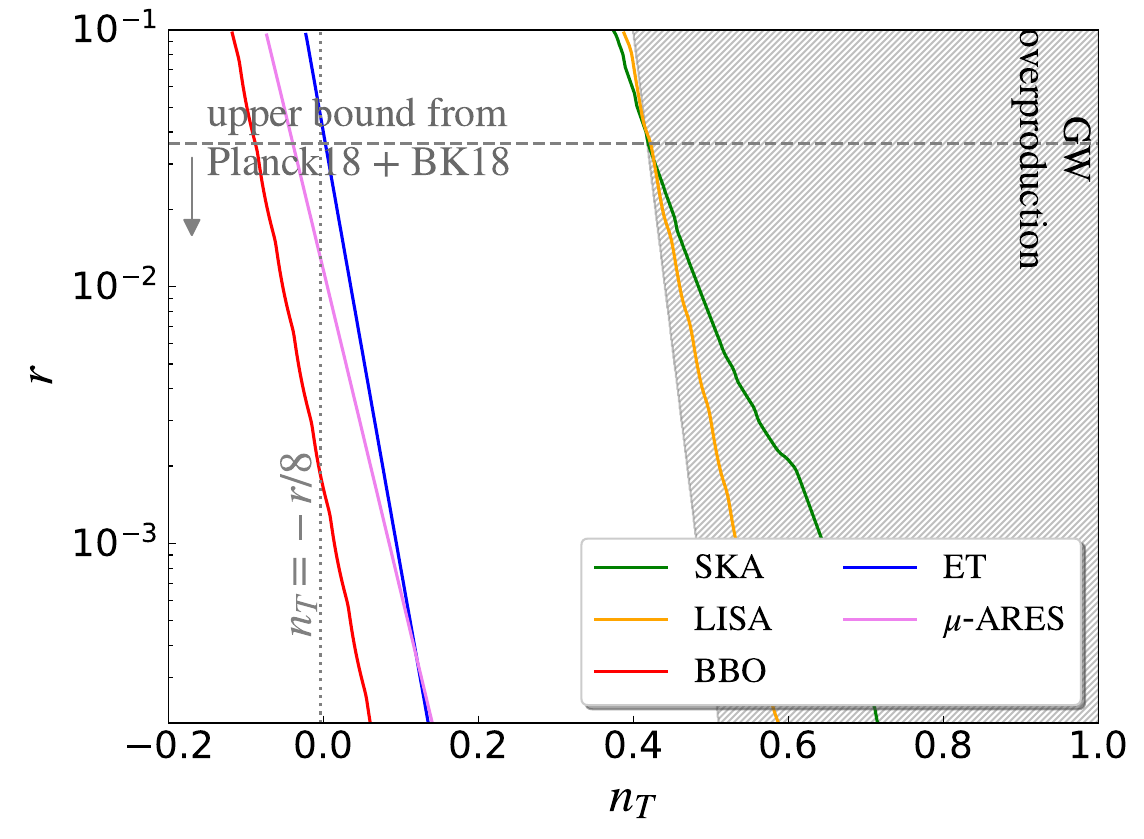}
    \includegraphics[scale=0.39]{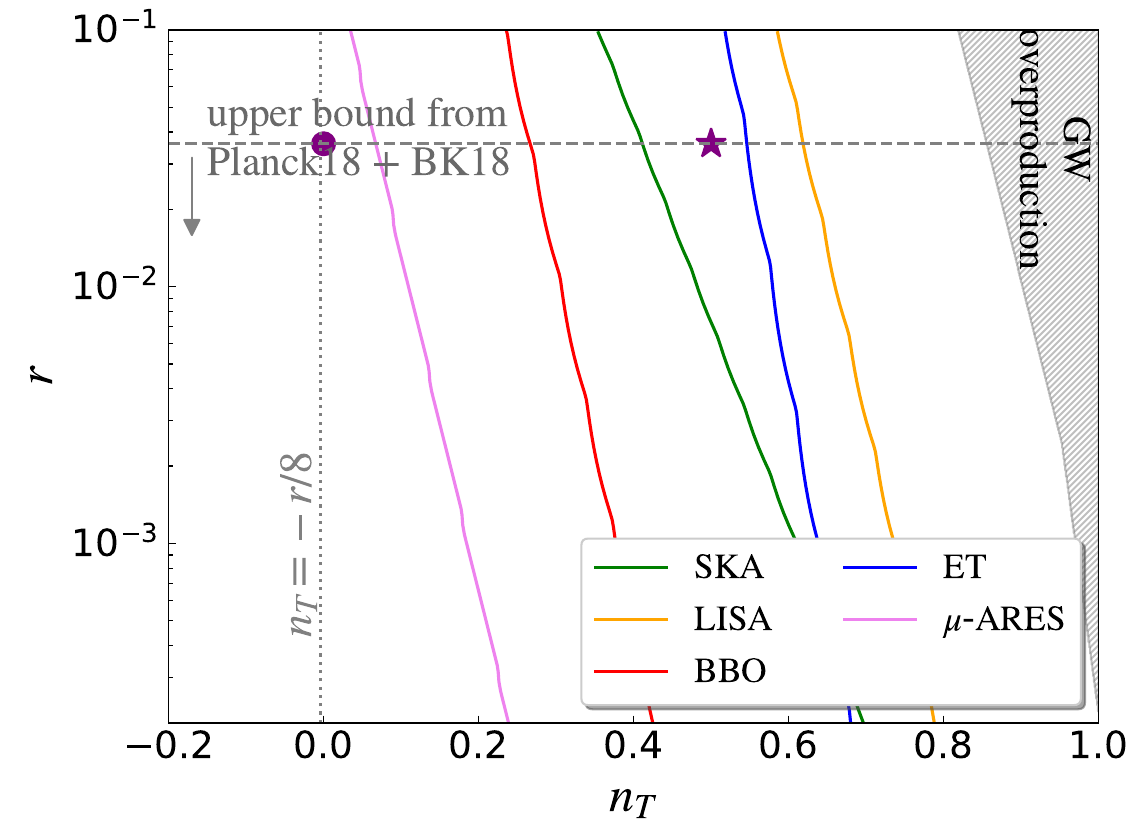}
    \caption[]{\justifying \it SNR in the $\nt-r$ plane across several GW detectors. \textbf{Left panel} corresponds to the standard scenario (without HQD) whereas \textbf{right panel} presents the HQD scenario for the same set of benchmark parameters ($m_Q=10^{14}$ GeV, $d=6$ and $\Lambda=m_{\rm Pl}$) as in Fig.~\ref{fig:GW_spectrum}. The purple star and circle in the right panel mark the benchmark point in the $\nt-r$ plane, which are also utilized in Fig.~\ref{fig:GW_spectrum}. Both panels assume $T_{\rm RH}=10^{16}$ GeV. Coloured lines correspond to $\rm SNR=10$ contour for the respective detector, with the region to the right $\rm SNR>10$. Hatched region presents the excluded region by Planck-18+BAO due to overproduction GWs.}
    \label{fig:ntr}
\end{figure*}

In Fig.~\ref{fig:ntr}, we show the regions of the $(\nt,r)$ parameter space where the SNR $>10$ for a range of proposed GW experiments both for absence and presence of HQD. Throughout this analysis, we assume a reheating temperature of $T_{\rm RH}=10^{16}$ GeV and use the same parameter as shown by the black solid line in Fig.~\ref{fig:GW_spectrum}, \textit{i.e.} $m_Q=10^{14}$ GeV, $d=6$ and $\Lambda=m_{\rm Pl}$, to present the HQD. We also mark $(\nt,r)=$ $(0, 0.036)$ and $(0.5,0.036)$ by purple circle and star in the right panel of the figure which are considered in Fig.~\ref{fig:GW_spectrum}. Each solid coloured line represents SNR $=10$ contour, corresponding to each GW detector, with the right hand portion of each contour indicates SNR $>10$. Although the SNR enhances as $\nt$ increases, higher $\nt$ values are excluded by Planck-18+BAO data, indicated by grey hatched region. However, compared to standard scenario (without HQD), presence of HQD allows higher values of $\nt$ (discussed in the previous section) as also shown the plots. The vertical dashed line at $\nt\simeq -r/8$ denotes the slow-roll consistency relation, which can be tested by BBO, ET, and $\mu$-ARES. In contrast, the detection prospects for LISA and SKA require a blue-tilted spectrum with $\nt>0$. Overall, the figure highlights that primordial GWs can be probed across a broad range of parameters. While enhanced detectability favours positive values of $\nt$, the slow-roll prediction remains encompassed within this parameter space. Moreover, any such signal is expected to be further modified by an early matter domination phase, driven by the presence of heavy quark, which constitutes the primary focus of our investigation.

\begin{table}[!ht]
    \centering
    \renewcommand{\arraystretch}{1.3}
    \begin{tabular}{|c|c|c|c|}
    \hline
    \hline
       Detectors  & Frequency range & $\tau_{\rm obs}$\\
    \hline
       SKA  & $\left[10^{-9}-4\times10^{-7}\right]$ Hz & $15$ years\\
       $\mu$-ARES & $\left[10^{-7}-1\right]$ Hz & $4$ years\\
       LISA  & $\left[10^{-4}-1\right]$ Hz & $4$ years\\
       BBO & $\left[10^{-3}-7\right]$ Hz & $4$ years\\
       ET  & $\left[1-10^3\right]$ Hz & $5$ years\\
    \hline
    \hline
    \end{tabular}
    \caption[]{\justifying \it Specification of the GW detectors in terms of operating frequency range and observational time period which is considered.}
    \label{tab:detector_spec}
\end{table}

\begin{figure*}
    \centering
    \includegraphics[scale=0.4]{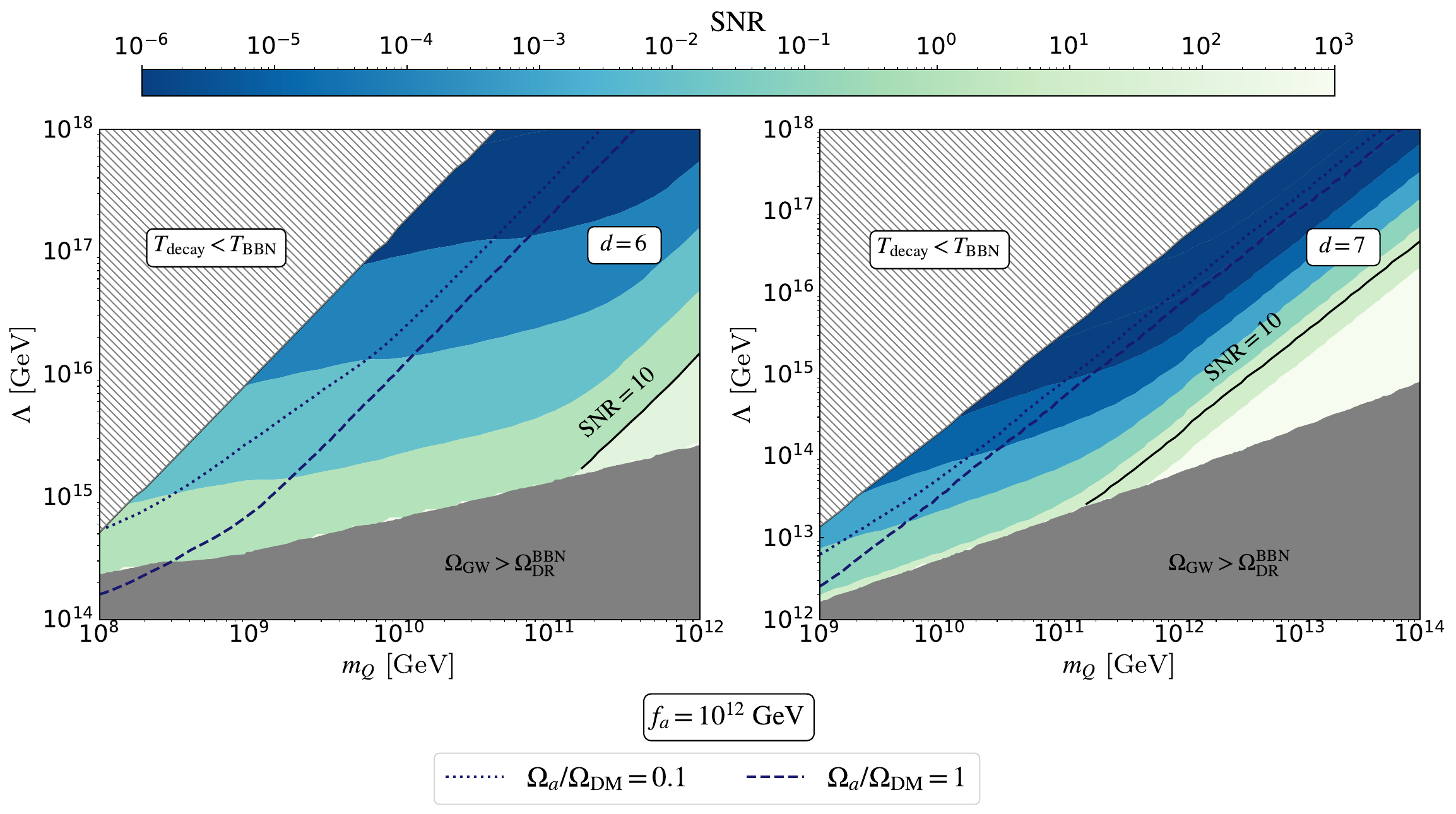}
    \caption[]{\justifying \it Illustration of SNR in the $m_Q-\Lambda$ plane for \textbf{LISA} across different dimensions. Black solid line represents ${\rm SNR}=10$, with the region below this line corresponding to ${\rm SNR} > 10$, as indicated by the colour scale. The gray-shaded region denotes the parameter space excluded by BBN constraints due to the overproduction of gravitational waves. Both the dimensions are for $\nt=0.5$ and $r=0.036$.}
    \label{fig:snr_lisa}
\end{figure*}

\begin{figure*}
    \centering
    \includegraphics[scale=0.4]{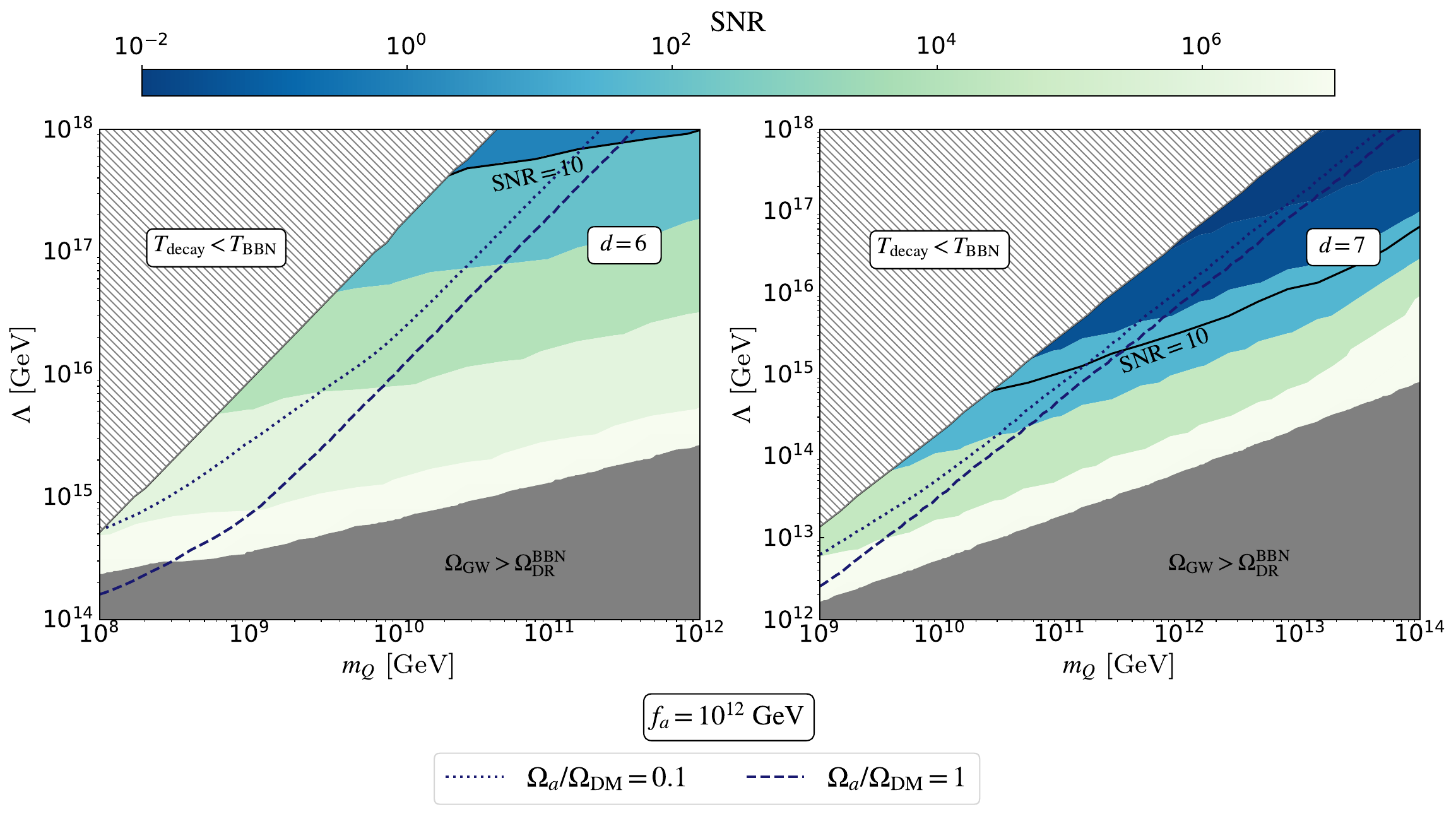}
    \caption[]{\justifying \it Descriptions of the figure is same as Fig.~\ref{fig:snr_lisa}, but for \textbf{ET}.}
    \label{fig:snr_et}
\end{figure*}

In Figs.~\ref{fig:snr_lisa}-\ref{fig:snr_et}, we have illustrated the SNR in $m_Q-\Lambda$ plane, considering $d=6$ and $d=7$, for LISA and ET, respectively (SNR plots for the other GW detectors are illustrated in the appendix~\ref{app:SNR}). The analysis is conducted for $\nt=0.5$ and $r=0.036$ to exhibit optimal sensitivity. The black and white hatched region come from BBN constraints as in section~\ref{subsec:axion_cosmo}. In each figure, the black solid lines represents SNR $=10$, with the region below these lines corresponding SNR $>10$. Our findings reveal that for a given detector and specific ($m_Q,\Lambda$) point, the SNR is generally higher for $d=6$, compared to $d=7$. This trend can be understood from the sensitivity plots which depicts that increasing the dimension strengthen the domination of early matter era, thereby reducing the detection prospect at the detectors (see, for instance, Fig.~\ref{fig:GW_spectrum}). 
Additionally, the plots indicate that SNR increases as $\Lambda$ decreases for fixed $m_Q$, consistently across both dimensions and all the detectors. However, the dependence on $m_Q$ at fixed $\Lambda$ exhibits opposite behaviour. This trend can be understood by remembering that duration of HQD which increases with increasing $\Lambda$ and decreasing $m_Q$. Consequently, achieving a reasonable SNR for each detector requires $\Lambda$ to be large when $m_Q$ is also large, a result that holds for both decay dimensions.
The dark gray-shaded region in the figures corresponds to GW overproduction as determined by Eq.~\eqref{eq:omega_delta_nu}. Since $d=6$ results in less damping of the GW spectrum compared to $d=7$, the overproduction occurs for larger values of $\Lambda$ for $d=6$. Due to the same reason, SNR is higher for $d=6$ compared to $d=7$, for a particular $m_Q$ and $\Lambda$. Additionally, we observed the BBO, ET and $\mu$-ARES can probe larger $\Lambda$-values.

Additionally in the figures, we present the contours for the fraction of axion DM ($\Omega_a/\Omega_{\rm DM}$) by dark blue dotted ($10\%$) and dashed ($100\%$) lines. The contours correspond to $f_a=10^{12}$ GeV and are shown to indicate how a given $f_a$ interacts with our figures. This is important when assessing the complementary information from axion haloscopes and interferometry missions.  For $d=7$ decay, GW missions are mostly sensitive for $m_Q$, while ET exhibits promising detection potential to probe the axion DM across a large range of DM mass for both dimensions.

\medskip

\section{Axion Haloscopes and GW detectors}
\label{sec:haloscope_GWs}
In the previous two sections we have shown how the effective axion model parameters $m_Q$, $\Lambda$, $f_a$ and the dimension of decay, $d$, can be inferred from axion haloscope and GW experiments separately. In this section we combine the information and explore the complementarity of the two.  

\subsection{Current experimental landscape}

\begin{figure*}[!ht]
    \centering
    \includegraphics[width=1.0\linewidth]{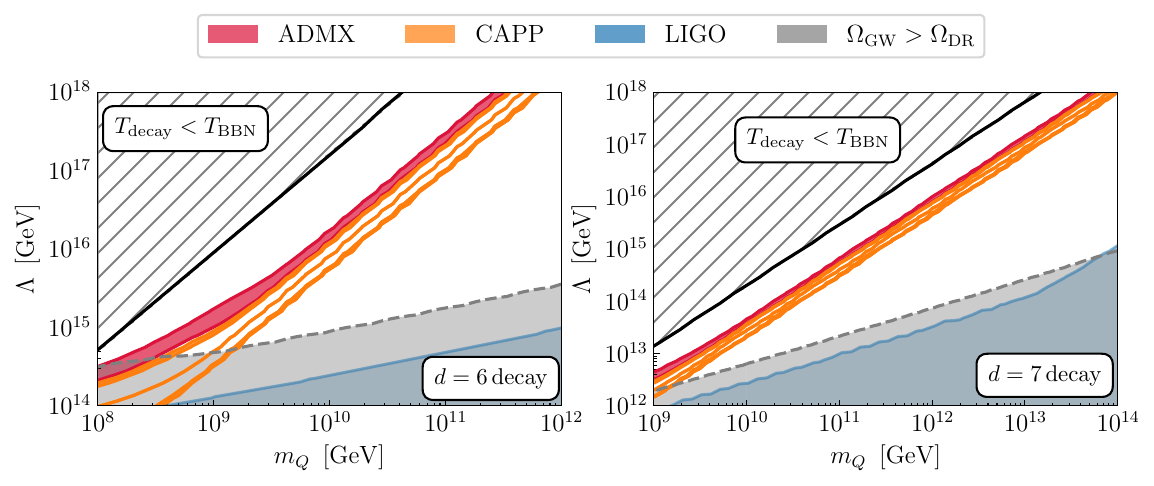}
    \caption[]{\justifying \it The same as Fig.~\ref{fig:current_experiments} but with the limits from GWs assuming $n_T=0.5$ and $r=0.036$. We show the limits from LIGO and the indirect bound on $\Omega_{\rm GW}$ coming from $\Delta N_{\rm eff}$. We use the result from Planck 2018 and BAO~\cite{Planck:2018vyg}.}
    \label{fig:current_GW}
\end{figure*}

As shown in Fig.~\ref{fig:current_experiments} and the surrounding discussion, the axion haloscope experiments CAPP and ADMX currently constrain small portions of the axion dark matter parameter space. Additionally, in section~\ref{sec:pgw}, we presented two existing experimental constraints that limit the parameter space for our axion models when assuming the optimistic inflationary GW background, $r=0.036$ and $n_T=0.5$. These are the results from Planck 2018 which limits the GW contribution to $N_{\rm eff}$ and LIGO which constrains $\Omega_{\rm GW}$ in the Hz frequency range. In Fig.~\ref{fig:current_GW} we present the additional GW bounds on the parameter space shown in Fig.~\ref{fig:current_experiments}. As exhibited in Figs.~\ref{fig:snr_ska}-\ref{fig:snr_et}, GW probes are insensitive to the breaking scale $f_a$ as well as whether or not the misalignment mechanism produce all to dark matter. However, by fixing $\Omega_a=\Omega_{\rm DM}$ we can determine, for a given point on the $\left(m_Q, \Lambda\right)$-plane, what value for $f_a$ is required. This in turn enables us to interpret whether a given axion haloscope experiment is sensitive to this point.  

By inspecting Fig.~\ref{fig:current_GW} we can see that, at least in the region of parameter space relevant for HQD, LIGO is less sensitive that constraints coming from Planck. We see in the right panel that LIGO becomes more constraining above $m_Q\sim 10^{14}\,{\rm GeV}$. The complementarity between experiments will become much more interesting in the near future because both axion haloscope experiments and GW detectors are set to improve substantially.

\subsection{Future experimental prospects}

\begin{figure*}[!ht]
    \centering
    \includegraphics[width=1.0\linewidth]{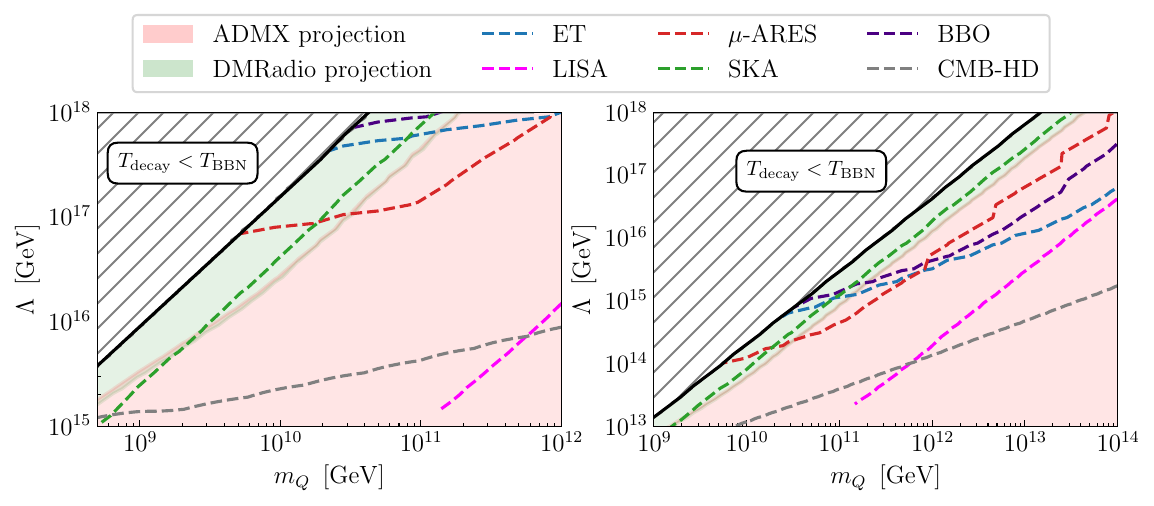}
    \caption[]{\justifying \it Future experimental landscape for axion dark matter and GW detectors. In the colored shaded regions we show the parameter regions where ADMX and DMradio are projected to be sensitive to. The dashed lines show the projected reach of future GW detectors assuming $n_T=0.5$ and $r=0.036$.}
    \label{fig:future_exps}
\end{figure*}

The future GW experiments we focus on in this work are discussed in detail in section~\ref{sec:pgw}. The prospects for axion haloscope experiments are similarly impressive. In particular the projections of DMRadio~\cite{DMRadio:2022pkf}, FLASH~\cite{Alesini:2023qed}, BabyIAXO-RADES~\cite{Ahyoune:2023gfw}, and ADMX~\cite{Stern:2016bbw} are projected to probe substantial portions of the $m_a$ range that reproduce the correct relic abundance with post-inflationary PQ breaking. In Fig.~\ref{fig:future_exps} we only highlight DMRadio and ADMX (green shaded and red shaded regions respectively) for clarity, the sensitivities of FLASH and BabyIAXO-RADES lie in the boundary area of the two experiments shown. Overlaid we show dashed lines which present the sensitivity projections for ET (blue), LISA (magenta), $\mu$-ARES (red), SKA (green), BBO (indigo) as presented in section~\ref{sec:pgw}. We also show the projected sensitivities coming from improved measurements of the CMB, which will give a constraint on the additional relativistic degrees of freedom, $\Delta N_{\rm eff}$. The most sensitive of which is reported to be CMB-HD~\cite{CMB-HD:2022bsz} (gray dashed line). We remind the reader that the SNR of each GW experiment below the SNR$=10$ line, therefore each experiment will be sensitive to parameters below the dashed lines. 

From Fig.~\ref{fig:future_exps} one can see that the DMRadio sensitivity region is closer to the $\tdec<T_{\rm BBN}$ region, this is because DMRadio will be sensitive to lower values of $m_a$ (high $f_a$). Therefore, the misalignment production needs to be suppressed to a greater extent, which can be achieved by minimizing $\tdec$. A later $\tdec$ corresponds to a suppression of $\Omega_{\rm GW}$ at lower $k$ as seen in Fig.~\ref{fig:GW_spectrum}. The wave-number associated with $T_{\rm BBN}$ is actually lower than the SKA frequency band, therefore for such low $m_a$ regions that DMRadio will be sensitive to, SKA is not able to probe. Alternatively, BBO, ET and LISA will be able to probe the DMRadio regions to a greater extent. This is because shorter HQD periods allow for the blue-tilted GW spectra to reach high enough amplitudes in higher frequency bands.

\begin{figure*}[!ht]
    \includegraphics[width=0.49\linewidth]{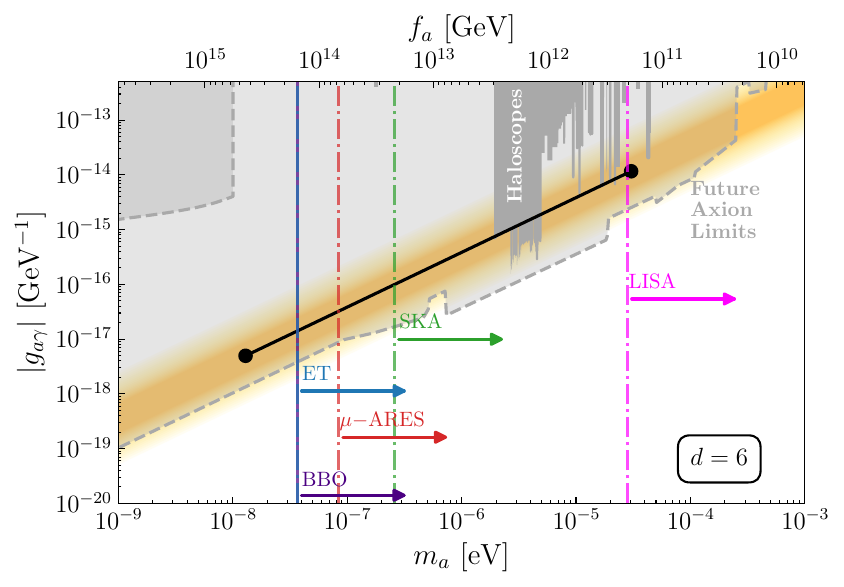}
    \includegraphics[width=0.49\linewidth]{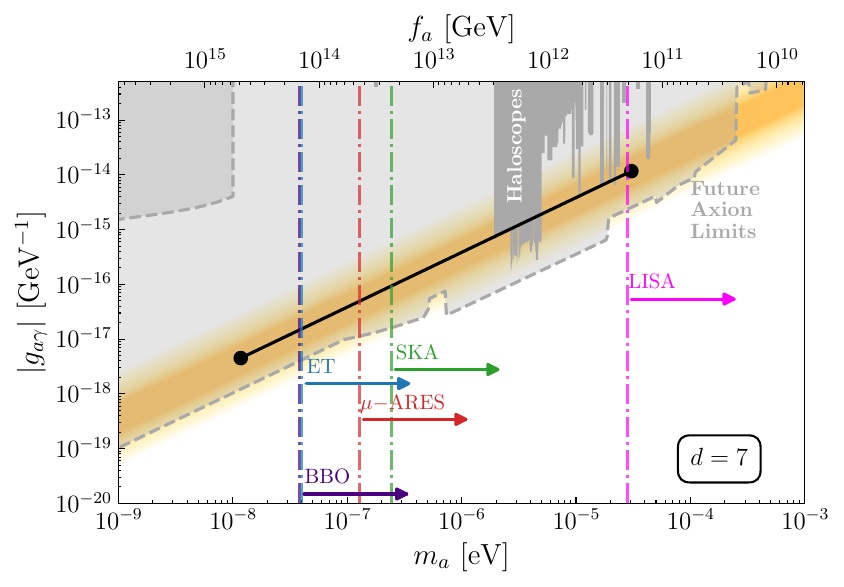}
    \caption[]{\justifying \it Similar to Fig.~\ref{fig:EMD_d6_d7} but with the possible reaches of future GW detectors. \textbf{Left panel} is for $d=6$ and \textbf{right panel} is $d=7$.}
    \label{fig:EMD_d6_d7_GW}
\end{figure*}

We emphasise that we have chosen the parameters $n_T$ and $r$ in order to see the potential GW experiments have to probe high-scale axion models. Therefore the GW lines are not to be read as future constraints on the axion model, they are instead maximum sensitivity lines. In that spirit, we take all the values in the $\left(m_Q, \Lambda\right)$-plane with ${\rm SNR}\geq 10$ for each experiment. From these points we determine the required $f_a$ to reproduce $\Omega_a=\Omega_{\rm DM}$. We take the maximum $f_a$ obtained and plot the result as a vertical line shown in Fig.~\ref{fig:EMD_d6_d7_GW}. We show the $\left(m_a, g_{a\gamma}\right)$-plane in order to make a stronger connection with axion experiments. We see that BBO, ET have the greatest potential, where the have maximum sensitivity to $f_a \sim 2\times10^{14}\,{\rm GeV}$ and $\mu$-ARES is not far behind $f_a\sim 7\times 10^{13}\,{\rm GeV}$, whereas SKA has the potential to probe below $f_a\approx2\times 10^{13}\,{\rm GeV}$ and $f_a\approx2\times 10^{13}\,{\rm GeV}$ for $d=6$. For models where quark decay proceeds via a $d=7$ effective operator, the maximum $f_a$ degrades slightly for ET, BBO and $\mu-$ARES but improves slightly for SKA. In both case LISA is essentially only sensitive to scenarios where HQD doesn't occur or when it doesn't alter the misalignment mechanism, $\tdec>T_{\rm osc}$. 

\medskip

\section{Discussion and Conclusion}
\label{sec:conclusion}

In this article we have explored the potential to use inflationary GWs to gain information about high-scale QCD axion models. Recent work has shown that many simple QCD axion models generate an early period of heavy quark domination, this has the consequence of increasing the number of viable post-inflationary axion models. Interestingly, the number of model-types which do not have a domain wall problem has now increased from 2 to 8. The non-standard cosmology however alters the misalignment mechanism requiring a larger $f_a$ (lower $m_a$) to reproduce the correct dark matter abundance.  

We have presented an effective model framework to analyse the axion models, focusing on heavy quark decays that proceed through dimension 6 and 7 terms. By varying all relevant parameters, $f_a,\,m_Q,$ and $\Lambda$, we show that there is substantial freedom in the mass of the QCD axion dark matter, this has not been pointed out previously. As a consequence, we have found the post-inflationary PQ breaking scenario can easily reproduce similar results than those typically associated with pre-inflationary breaking, i.e. $m_a\lesssim10^{-5}\,{\rm eV}$ and $\Delta N_{\rm eff}\ll0.027$.

We propose using the stochastic GW background generated from inflation as a way to determine whether a period of HQD occurred. By using such GWs to trace the early epochs of the Universe, we can obtain information on the particle physics that leads to matter domination. We investigated $\nt=0$ spectrum and its detectability with the GW detectors and then to assess the maximum sensitivity these probes will have, we optimistically take a blue-titled GW power spectrum, $n_T=0.5$ in Fig.~\ref{fig:GW_spectrum} for illustration purpose only. Such spectra are well motivated by a range of alternatives to the standard single field, slow-roll inflation, to list a few of them~\cite{Liddle:1993fq,Brandenberger:2006xi,Calcagni:2013lya,Fujita:2018ehq,Cook:2011hg,Mukohyama:2014gba,Anber:2009ua,Baldi:2005gk,Oikonomou:2021kql,Biagetti:2013kwa,Caldwell:2017chz,Dimastrogiovanni:2018xnn,Adshead:2017hnc,Kuroyanagi:2020sfw,NANOGrav:2023hvm}. A more comprehensive study would allow the parameters that control the GW power spectrum $A_T$ and $n_T$, to vary, however by setting optimistic but valid values for such parameters we effectively determine the possible reach of future interferometers. 

We found that GWs are only sensitive to two parameters in our effective setup, $m_Q$ and $\Lambda$, as well as dimension of decay, $d$. In all cases large portions of the particle physics parameter space could be probed in future GW experiments, see Fig.~\ref{fig:EMD_d6_d7_GW}. When fixing the QCD axion to be 100\% of the dark matter observed we can then fix the decay constant $f_a$ allowing us to understand the complementarity between GW probes and axion haloscope experiments. We highlight this in Fig.~\ref{fig:EMD_d6_d7_GW} where we show BBO and ET exhibit the highest potential, capable of probing scenarios where $m_a\gtrsim 3\times 10^{-8}$ eV, and $\mu-$ARES can reach $m_a>10^{-7}$ eV, for dimension $6$ and $7$ decays. Notably, the difference between dimension $6$ and $7$ decay for probing axion mass is not significantly large.

If the characteristic features of the GW spectral shapes proposed in this study are observed, one may then perform parameter reconstructions for the effective models we investigate. With only the information of the GW spectrum, distinguishing between HQD and other forms of early matter domination may prove challenging. It is only with the additional confirmation from axion haloscope experiments will the HQD explanation look more likely than other forms of early matter domination. At this point, by measuring the strength of $g_{a \gamma}$ one may be able to infer the anomaly coefficients and therefore the corresponding axion models. 

On the other hand, in the event of a positive detection of a light QCD axion, no $\Delta N_{\rm eff}$ and no of inflationary GWs, HQD would remain as plausible as pre-inflationary case. This is because our investigation has relied on the blue-tilt of the spectrum. It is true that a slightly red-tilted spectrum may still be detectable in future GW detectore. We believe that further investigations into experimental methods to probe the possibility that HQD occurred in the early Universe are warranted. 
GAN~\cite{McAllister:2017lkb}, RADES~\cite{Melcon:2018dba}, and QUAX~\cite{Alesini:2020vny}. Experiments using a different technology but probing a similar mass range are MADMAX~\cite{MADMAX:2019pub}, BABYIAXO~\cite{Ahyoune:2023gfw} and FLASH~\cite{Alesini:2023qed}. For lower axion masses in the $10^{-9}$ to $10^{-7}$ eV range, experiments such as SRF~\cite{Berlin:2020vrk, Tang:2023oid}, DMRadio~\cite{DMRadio:2022pkf}, and ABRACADABRA~\cite{Salemi:2021gck} show particular promise.

In this paper we have highlighted that the precision era of gravitational wave cosmology holds significant potential for meaningfully probing fundamental physics. We have presented a proof-of-concept whereby the measurement of inflationary gravitational waves could break the experimental deadlock between pre- and post-inflationary PQ breaking scenarios for dark matter axion models.

\medskip

\section*{Acknowledgement}
AC is supported by the National Natural Science Foundation of China (NSFC) through the grants No.12425506, 12375101, 12090060, and 12090064, and the SJTU Double First Class start-up fund (WF220442604). DP thanks Indian Statistical Institute (ISI), Kolkata for financial support through Senior Research Fellowship and the computational facilities of Technology Innovation Hub, ISI Kolkata. The authors also thank Supratik Pal, Lucien Heurtier and Enrico Nardi for fruitful discussions and Leszek Roszkowski for comments of the manuscript.


\appendix
\section{Noise modelling for the GW detectors}
\label{app:noise}
As already mentioned in Sec.~\ref{subsec:snr}, estimate of noise spectra associated with the detectors, while assessing the signal. In our study, to asses the signal, we consider the instrumental noise for the aforementioned GW detectors. For LISA and BBO, noise spectra is estimated in terms of effective noise power spectral density ($S(f)$), which can be related with the noise spectra of the detectors as~\cite{Gowling:2021gcy}:
\begin{eqnarray}\label{eq:omegagw}
    \Omega_{\rm GW}^{\rm noise} (f) \;=\; \left(\frac{4\pi^2}{3H_0^2}\right)f^3 S(f).
\end{eqnarray}

\begin{itemize}
    \item \textbf{LISA:} Instrumental noise of LISA has two key sources: acceleration noise of test mass (acc) and optical metrology noise (omn)~\cite{Gowling:2021gcy}. In terms of power spectral density, instrumental noise is expressed as~\cite{Breitbach:2018ddu}: 
\begin{align}
S_{\rm acc}(f) &= 9 \times 10^{-30} \frac{1}{(2\pi f / 1{\rm Hz})^4} \left( 1 + \frac{10^{-4}}{f / 1{\rm Hz}} \right)~{\rm m^2Hz^{-1}},\\
S_{\rm oms}(f) &= \left(1.5\times 10^{-11} {\rm m}\right)^2\left[1+\left(\frac{2 {\rm mHz}}{f}\right)^4\right]\,{\rm Hz}^{-1}.
\end{align}
Thus, we have for LISA 
\begin{eqnarray}
S_{\rm LISA}(f) &= \frac{10}{3L^2} \left(2S_{\rm acc}(f)\left(1+{\rm cos}^2 \left(\frac{f}{c/(2\pi L)}\right)\right)\right.\nonumber \\
&\left.+ S_{\rm oms}(f)\right) \left[1 + \frac{6}{10}\left( \frac{f}{c/(2\pi L)} \right)^2\right],
\label{eq:SeffLISA}
\end{eqnarray}
where $L = 2.5 \times 10^9$~m is the arm-length of LISA. Thus, using Eq.~\eqref{eq:omegagw}, the GW energy density power spectrum can be calculated.

    \item \textbf{BBO:} For BBO, the non sky-averaged instrumental noise power spectral density is~\cite{Yagi:2011yu} 
\begin{eqnarray}
S_{\rm BBO}(f) &= \left[1.8 \times 10^{-49} (f / 1{\rm Hz})^2+2.9 \times 10^{-49}\right. \nonumber\\ 
& \left.+ 9.2 \times 10^{-52} (f / 1{\rm Hz})^{-4}\right]~{\rm Hz^{-1}}.
\label{eq:SeffBBO}
\end{eqnarray}
Hence, using Eq.~\eqref{eq:omegagw}, $\Omega_{\rm GW}^{\rm noise}$ for BBO can be calculated.

\end{itemize}

For \textbf{ET} and \textbf{SKA}, we have followed Ref.~\cite{Schmitz:2020syl} to generate the noise spectra. The characteristic strain of noise ($h_c^{\rm noise}$) for the instrumentation of $\mathbf{\mu}$\textbf{-ARES} is provided in Ref.~\cite{Sesana:2019vho}. The noise spectra can be calculated using $h_c$ as~\cite{Maggiore:1999vm}
\begin{eqnarray}
    h_c^{\rm noise}(f)\simeq1.26\times^{-18}\left(\frac{1\,{\rm Hz}}{f}\right)\sqrt{h^2 \Omega^{\rm noise}_{\rm GW}(f)}.
\end{eqnarray}

\section{Accessing the GW signal for SKA, \texorpdfstring{$\mu$}{mu}-ARES and BBO}
\label{app:SNR}
\begin{figure*}
    \centering
    \includegraphics[scale=0.4]{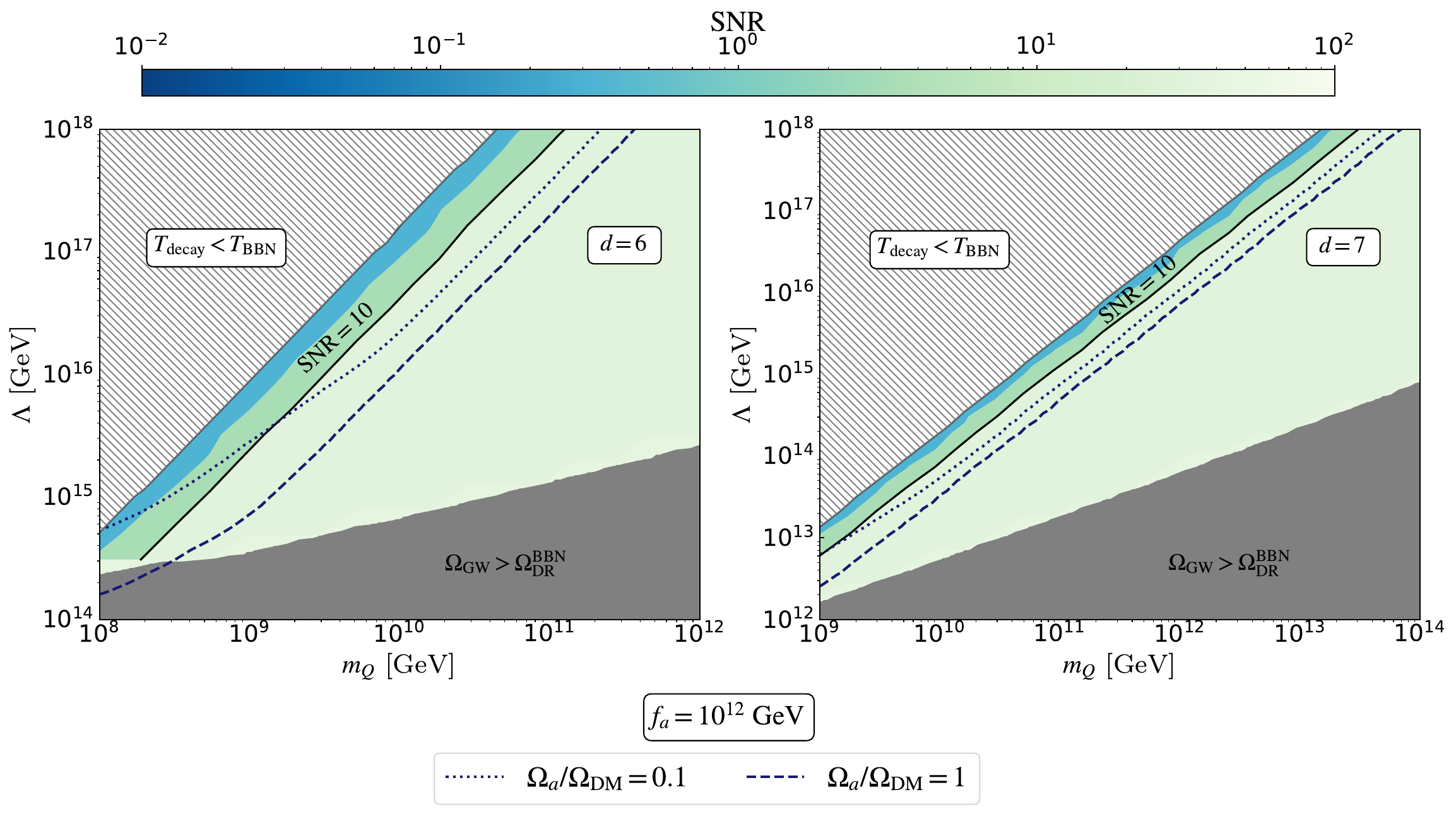}
    \caption[]{\justifying \it Illustration of SNR in the $m_Q-\Lambda$ plane for \textbf{SKA} across different dimensions. Black solid line represents ${\rm SNR}=10$, with the region below this line corresponding to ${\rm SNR} > 10$, as indicated by the colour scale. The gray-shaded region denotes the parameter space excluded by BBN constraints due to the overproduction of gravitational waves. Both the dimensions are for $\nt=0.5$ and $r=0.036$.}
    \label{fig:snr_ska}
\end{figure*}

\begin{figure*}
    \centering
    \includegraphics[scale=0.4]{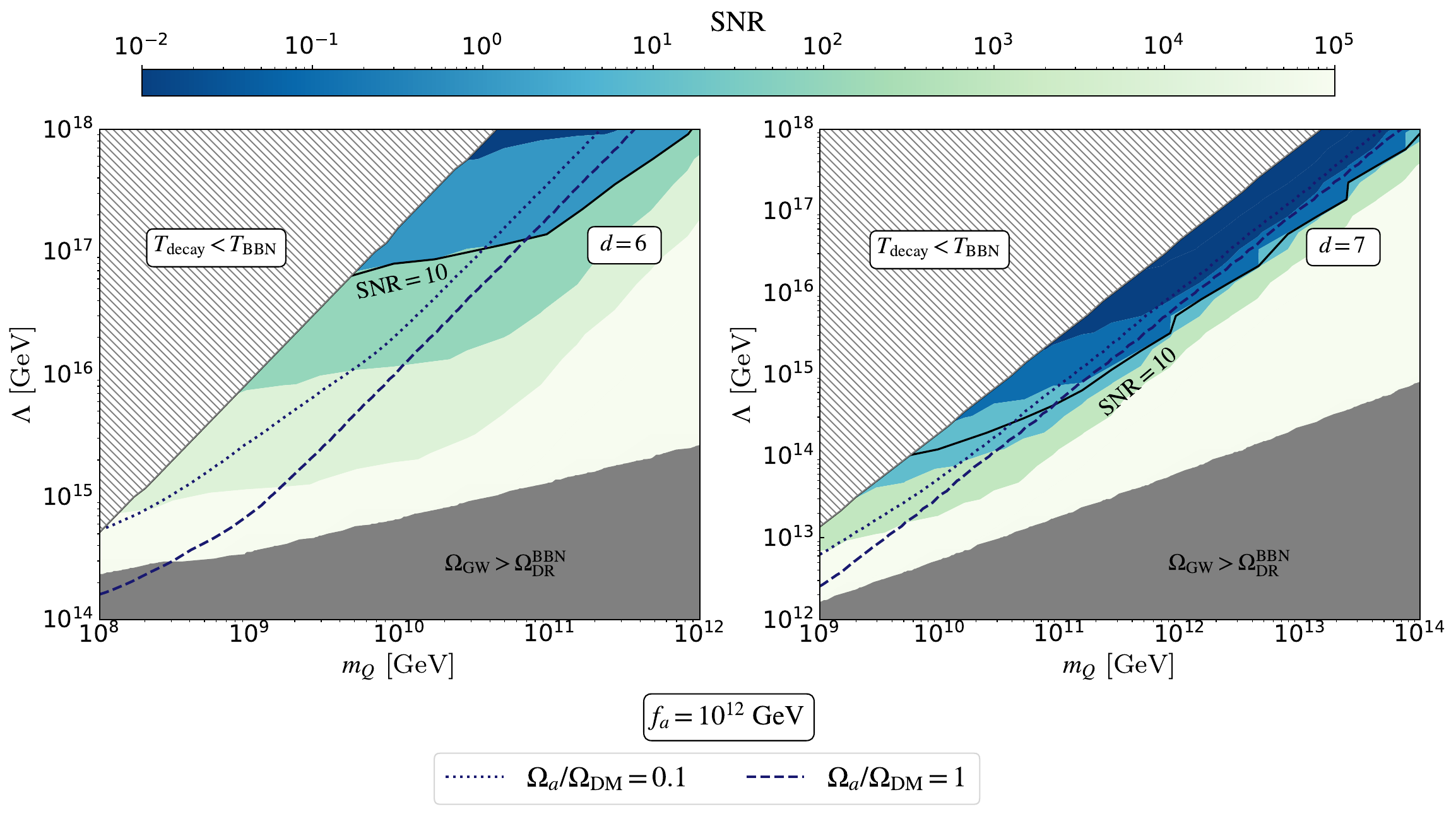}
    \caption[]{\justifying \it Descriptions of the figure is same as Fig.~\ref{fig:snr_ska}, but for $\mathbf{\mu}$\textbf{-ARES}.}
    \label{fig:snr_muares}
\end{figure*}

\begin{figure*}
    \centering
    \includegraphics[scale=0.4]{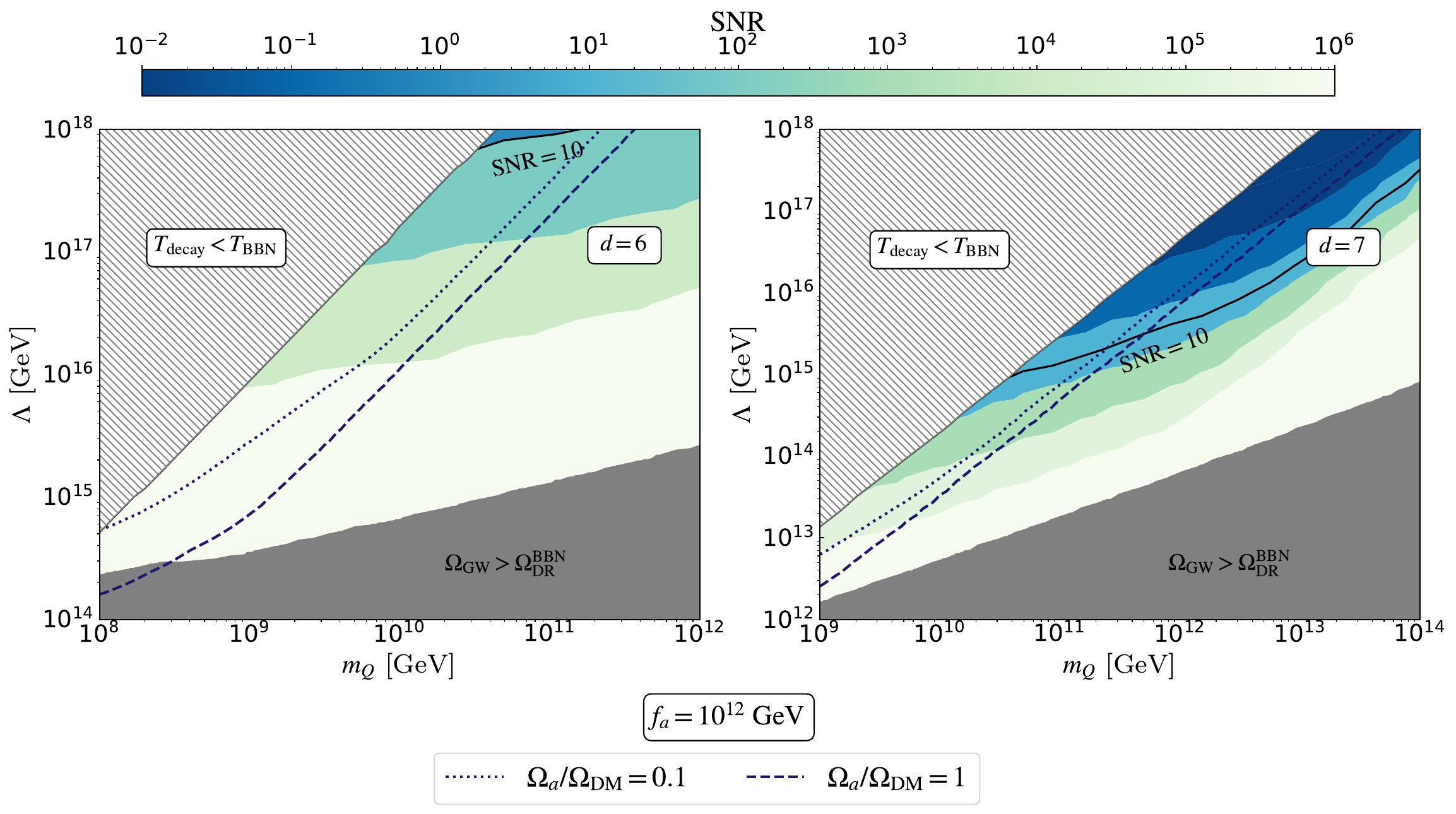}
    \caption[]{\justifying \it Descriptions of the figure is same as Fig.~\ref{fig:snr_ska}, but for \textbf{BBO}.}
    \label{fig:snr_bbo}
\end{figure*}

Figs.~\ref{fig:snr_ska}-\ref{fig:snr_bbo} show the SNR in $m_Q-\Lambda$ plane for $d=6$ and $d=7$, respectively for SKA, $\mu$-ARES and BBO. As in the LISA and ET analysis in Sec.~\ref{subsec:snr}, the results are presented for $\nt=0.5$ and $r=0.36$, considering instrumental noise of each detector. In all figures, the black solid curve denotes SNR$=10$, with the region below this curve corresponding to SNR$>10$. As discussed in Sec.~\ref{subsec:snr}, the case $d=6$ yields larger SNR than $d=7$ across all the GW detectors. Moreover, the SNR is directly proportional to $m_Q$ for a fixed $\Lambda$, however, it is inversely proportional to $\Lambda$ for a fixed $m_Q$ (discussed in detailed in Sec~\ref{subsec:snr}). The dark blue dotted and dashed curves represent the axion DM fraction $10\%$ and $100\%$, respectively. Thus, the detectors exhibit promising regions in the $m_Q-\Lambda$ plane where axion DM can probed over a broad parameter range.

\newpage
\bibliography{ref}
\end{document}